\documentclass[natbib]{svjour3}        

\usepackage[dvips]{epsfig}
\usepackage[latin1]{inputenc}
\usepackage{graphicx,amssymb,mathrsfs,amsmath,amsfonts,color,psfrag}
\newcounter{myenum}

\renewcommand{\figurename}{Figure}
\renewcommand{\d}{\mbox{d}}

\newcommand{\figref}[1]{\figurename~\ref{#1}}
\newcommand{\secref}[1]{Section~\ref{#1}}
\newcommand{\ud}{\mathrm{d}}
\newcommand{\bo}[1]{\boldsymbol{#1}}
\newenvironment{flushitemize}{%
  \begin{list}{$\diamond$}%
    {\setlength{\leftmargin}{.6cm}%
     \setlength{\labelwidth}{0.15cm} 
     \setlength{\itemindent}{0em} 
     \setlength{\labelsep}{0.4cm}}
     \usecounter{myenum}}%
  {\end{list}}
\newenvironment{flushenumerate}{%
  \begin{list}{(\arabic{myenum})}%
    {\setlength{\leftmargin}{0pt}}%
     \setlength{\labelwidth}{0pt} 
     \setlength{\itemindent}{0.5em} 
     \setlength{\labelsep}{0.5em} 
     \usecounter{myenum}}%
  {\end{list}}  

\journalname{Bulletin of Mathematical Biology}

\begin{document}

\title{Diffusion of finite-size particles in confined geometries}
\author{Maria Bruna  \and S. Jonathan Chapman}

\institute{Mathematical Institute, University of Oxford, 24-29 St. Giles', Oxford OX1 3LB, United Kingdom \\
              \email{bruna@maths.ox.ac.uk \and chapman@maths.ox.ac.uk}
}
\date{Received: 13 December 2012 / Accepted: 22 April 2013}


\setcounter{secnumdepth}{3}
\setcounter{tocdepth}{2}

\maketitle

\begin{abstract}
The diffusion of finite-size hard-core interacting particles in two- or three-dimensional confined domains is considered in the limit that the confinement dimensions become comparable to the particle's dimensions.  The result is a nonlinear diffusion equation for the one-particle probability density function, with an overall collective diffusion that depends on both the excluded-volume and the narrow confinement.  By including both these effects the equation is able to interpolate between severe confinement (for example, single-file diffusion) and unconfined diffusion. Numerical solutions of both the effective nonlinear diffusion equation and the stochastic particle system are presented and compared.  As an application, the case of diffusion under a ratchet potential is considered, and the change in transport properties due to excluded-volume and confinement effects is examined.

\end{abstract}

\section{Introduction} \label{intro}

Transport of material under confined conditions occurs throughout nature and  applications in industry. Examples include the transport of particles in biological cells, such as ion channels that conduct ions across the cell surface \citep{Hille:2001tw} or intracellular cargo along microtubule filaments \citep{alberts2002molecular,Klumpp:2005ed}, and in zeolites \citep{Keil:2000ev}. Similarly, confinement can be important in the diffusion of cells themselves (e.g. blood cells through microvessels, \citealp{Pries:1996ik}) and surface diffusion on the cell membrane, which is usually crowded with fixed and mobile obstacles  \citep{NicolauJr:2007dy}. Moreover, recent advances in nanotechnology have allowed the development of synthetic nanopores and microfluidic devices \citep{Hanggi:2009ha}, which can be used for the sensing of single particles (such as small molecules, organic polymers, proteins, or enzymes) and for studying chemical reactions, biomolecular recognition, and interactions at the nanoscale \citep{Dekker:2007kc, Howorka:2009iw}. 
A common feature in these applications is the interplay between the particle motion (usually noisy) and the geometric constraints. An additional factor comes into play if the system contains a collective of  interacting particles rather than an individual particle. The way in which these characteristics combine to produce the global behaviour is a crucial factor in the understanding of such systems. In many respects, these transport phenomena can be studied in terms of the canonical problem of geometrically constrained Brownian dynamics \citep{Burada:2009hr}.

When considering a theoretical model of particle diffusion in confined
environments, there are three important modelling decisions to
make. First, one must decide on the most appropriate representation of
the particle diffusion and interactions (with other particles and the
confining walls). For example, a common approach is to use a
lattice-based random walk model with exclusion \citep{Plank:2012fa},
that is, to assume that the motion of particles is restricted to
taking place on a lattice and that any attempted move to an occupied
site is aborted. An alternative approach is to consider a lattice-free
random walk, in which the individual particle movements are not
restricted to a lattice. It this case, excluded-volume interactions
can be taken into account by assuming particles are hard spheres which
cannot overlap each other, thus considering a Brownian motion of hard
spheres \citep{Bruna:2012cg}. While in some cases a lattice-based
model is more suitable for the particular application, in general the
lattice-free approach is more realistic \citep{Plank:2012fa} and the
choice of an on-lattice model is for technical convenience only.  

The second modelling decision concerns the level of description, that
is, whether to use 
an individual-based model or a population-based model. In the
first case, the system of diffusing and interacting particles is
represented with a stochastic model that describe the dynamics and
interactions of each particle explicitly. This is typically a
computationally intensive approach, involving many statistically
identical realisations of the stochastic simulation to develop insight
into the population-level dynamics.  
In contrast, the population-based model consists of a continuum
description of the system in the form of a partial differential
equation (PDE) for the population density of individuals. The
continuum model tends to be easier to solve and analyse and can be
particularly useful when, for large systems of interacting particles,
discrete models become computationally intractable. However, the
challenge is to predict the correct PDE description of a given system
of interacting particles, and, as a result, many population-based
models are described phenomenologically at the continuum level rather
than derived from the underlying particle transport process. For
example, while it is well-understood that a non-interacting Brownian
motion is associated a linear diffusion PDE at the population-level,
it is not so straightforward to predict how excluded-volume
interactions at the discrete level emerge in the PDE model. As pointed
out in \cite{Plank:2012fa}, the ability to represent mathematically
both the individual-level details and the population-level description
of a stochastic particle system is important because many experimental
observations involve data at both levels for the same system.  As a
result, if we are to use both the individual-based and the
population-based models of the same system, the link between the two
must be fully understood to ensure that both models are consistent
with each other.  

Finally, the third consideration has to do with the way confinement is
included in the model. It is important to note that the idea of
confinement is inevitably relative the particle's characteristic
size.  A common approximation of confinement situations which is
applicable when the particles are much smaller than the channel width
is to ignore steric interactions between particles (assuming they are
simply points) and only consider the geometric effects of the
confining environment \citep{Burada:2009hr}. For example, the
diffusion of point particles in a (narrow) tube of varying
cross-section can be approximated by an effective one-dimensional
diffusion equation  known as the Fick--Jacobs equation
\citep{jacobs1967diffusion,Reguera:2001ev}. Another example in which
particle interactions are omitted can be found in the Brownian ratchet
models of molecular motors
\citep{MunozGutierrez:2012cz,Eichhorn:2002du}, which take the form of
a one-dimensional diffusion under a periodic potential and tilting
force.   

The opposite limit is  \emph{single-file diffusion}
\citep{Henle:2008fv}, in which the finite-size of particles is taken
into account but the confinement is so extreme that particles cannot
diffuse past each other (imagine a channel of width equal to the
diameter of particles). Mathematically this problem is modelled as a
one-dimensional domain with hard-core interacting particles (hard
rods) and has been widely studied (see, for example,
\citealp{Lizana:2009ic,Bodnar:2005kv}).  

Both of these limits are extremes. The distinguished limit in which
the finite-size interactions are important but the confinement is not
so extreme that particles cannot pass one another has received
little attention; one notable exception is the exclusion process on a lattice in \cite{Henle:2008fv}.

\subsection{Aim of this paper}

This work introduces a theoretical framework for studying particle diffusion processes in confined environments. Rather than attempting to answer a particular question related to one of the applications presented earlier, here we are interested in developing a technique to tackle the common first steps in any of these such problems. Following the three
considerations outlined above, we are interested in a lattice-free
approach, in deriving the population-level model \emph{systematically}
from the individual-based model, and in an intermediate level of
confinement. 
To this end, we consider the evolution of a system of $N$ identical
hard spheres in a confined domain, in the limit that the confinement
dimensions become comparable to the particle dimensions. In this
setting, the finite size of particles is important not only for 
particle--particle interactions, but also for  interactions with
the domain walls.  We consider in particular three confinement
scenarios: a two-dimensional channel, a three-dimensional square
channel, and two close parallel plates. However, since our approach is
systematic, our model can  be extended to other geometries.  

The key idea is that the system will reach equilibrium in the confined
directions quickly, leading to an effective diffusion of reduced
dimension in the unconfined directions only.
 With this in mind, the solution procedure
consists of two steps: first, to reduce the model of $N$ interacting
particles to a model for the evolution of the one-particle marginal
density, as we did in \cite{Bruna:2012cg}; and second, to reduce the
resulting model from a $d$-dimensional confined domain to an effective
one-dimensional axial model in the case of a narrow channel, or to an
effective two-dimensional planar model in the case of  parallel
plates.

\subsection{Plan of this paper}

The work is organised as follows: in the next section we will
introduce the problem setup, illustrate how the problem simplifies in
the case of point particles and present the main result of this work,
a population-level PDE model for the diffusion of hard spheres as a
function of a confinement parameter, given by equation
\eqref{fp_reduced}. In the third section we examine how our model
interpolates between the different limiting cases of confinement. In
Section 4 we explore numerical solutions of our PDE model and compare
them with stochastic simulations of the particle-based model and
numerical solutions of the limiting models. Finally, the fifth section
will be devoted to the derivation of \eqref{fp_reduced} for a
two-dimensional channel.  

\section{The model} \label{sec:model}

\subsection{The setup: drift-diffusion in confined geometries}
We consider a population of $N$ identical particles diffusing in a
bounded domain $\Omega \subset \mathbb R^d$ ($d=2,3$), interacting
with each other and the domain walls with a repulsive hard-core
potential, and in the presence of an external force. We work in the
dimensionless problem by scaling space with a typical unconfined
dimension $L$, time with $L^2/D_0$ where $D_0$ is the constant
molecular diffusion coefficient, and force with $\gamma D_0/L$ where
$\gamma$ is the frictional drag coefficient.
We assume particles are spherical with nondimensional
diameter $\epsilon\ll1$. 

Assuming the overdamped limit, the stochastic dynamics of the system is described by a set of stochastic Langevin equations
\begin{equation} 
\label{sde}
\d{\bf X}_i(t) =  {\bf f} \bo ({\bf X}_i(t) \bo ) \d t + \sqrt{2} \,
\d {\bf W}_i(t), \qquad i = 1, \dots N, 
\end{equation}
where ${\bf X} _i(t) \in \Omega$ denotes the centre of  particle $i$
at time $t\ge0$, $\bf f$ is the dimensionless external force (or
drift) and ${\bf W}_i$ are $N$ independent $d$-dimensional standard
Brownian motions. We note that by writing $ {\bf f} \bo ({\bf X}_i(t)
\bo ) $ we are assuming that the force acting on the $i$th particle
only depends on its own position, thus excluding forces such as the
electromagnetic force which would depend on the positions of all the
particles $\vec X=  ({\bf  X}_1, \ldots,  {\bf  X}_N)$. We suppose
that the initial positions ${\bf  X}_i(0)$ are random and identically
distributed. Note that, because of the finite size of particles,
we have the set of constraints $\|{\bf X}_i - {\bf X}_j \| \ge
\epsilon$ for $i \ne j$,  so that  
the system of SDEs \eqref{sde} is coupled.

The Langevin system \eqref{sde} is equivalent to the Fokker--Planck
equation for the joint probability density $P ({\vec x}, t)$ of the
$N$ particles to be found at the position $\vec x = ({\bf x}_1, \dots,
{\bf x}_N)\in \Omega^N$ at time $t$, given by  
\begin{subequations}
\label{fp}
\begin{align}
\label{fp_eq}
\frac{\partial P}{\partial t} (\vec x, t) &= \nabla_{\vec x} \cdot \left [  \nabla_{\vec x} P - \vec F (\vec x) P \right ],
\end{align}
where $\vec \nabla _{\vec x}$ and $\vec \nabla _{\vec x} \, \cdot$
respectively stand for the gradient and divergence operators with
respect to the $N$-particle position vector $\vec x$ and $\vec F (\vec
X) = \bo ( {\bf f}({\bf  x}_1), \dots, {\bf f}({\bf  x}_N) \bo )$ is the total
drift vector. Because of excluded-volume effects, the domain of
definition of \eqref{fp} (or configuration space) is not $\Omega^N$
but its hollow form $\Omega_\epsilon ^N = \Omega ^N \setminus \mathcal
B_\epsilon$, where $\mathcal B_\epsilon=\{\vec x\in \Omega^N: \exists
i\ne j \textrm{ such that } \|{\bf  x}_i - {\bf  x}_j\| \le \epsilon
\}$ is the set of all illegal configurations (with at least one
overlap). On the contact surfaces $\partial \Omega_\epsilon^N$ we have
the reflecting boundary condition 
\begin{align}
\label{fp_bc}
0 &= \big[ \nabla_{\vec x} \,  P -  \vec F(\vec x) \, P \big] \cdot  {\vec n},
\end{align}
where $ {\vec n}  \in \mathcal S^{dN-1}$ denotes the unit outward
normal. Finally, since the particles are initially identically
distributed, the initial probability density $P(\vec x, 0) = P_0(\vec
x)$ is invariant to permutations of the particle labels. The form of
\eqref{fp} then means that $P$ itself is invariant to permutations of
the particle labels for all times. 
\end{subequations}

We suppose that $\Omega$ is a confined domain, with $k<d$
\emph{confinement} dimensions which are comparable to $\epsilon$. We
introduce $d_e = d-k$ as the \emph{effective} dimensionality of the
problem. In particular, we shall consider the following cases: 
\begin{subequations}
\label{Omeganarrow}
\begin{flushitemize}
\item \textbf{(NC2)} Two-dimensional narrow channel ($d=2$, $k=1$ and $d_e = 1$):
\begin{equation}
\label{OmegaNC2}
\Omega = \left[-\tfrac{1}{2}, \tfrac{1}{2}\right] \times \left[-\tfrac{H}{2}, \tfrac{H}{2}\right].
\end{equation}
\item \textbf{(NC3)} Three-dimensional narrow channel ($d=3$,  $k=2$ and $d_e =1$):
\begin{equation}
\label{OmegaNC3}
\Omega = \left[-\tfrac{1}{2}, \tfrac{1}{2}\right] \times \left[-\tfrac{H}{2}, \tfrac{H}{2}\right]\times \left[-\tfrac{H}{2}, \tfrac{H}{2}\right].
\end{equation}
\item \textbf{(PP)} Two parallel plates ($d=3$, $k=1$ and $d_e = 2$):
\begin{equation}
\label{OmegaPP}
\Omega = \left[-\tfrac{1}{2}, \tfrac{1}{2}\right] \times \left[-\tfrac{1}{2}, \tfrac{1}{2}\right] \times \left[-\tfrac{H}{2}, \tfrac{H}{2}\right],
\end{equation}
\end{flushitemize}
\end{subequations}
where $H = \mathcal O(\epsilon)$ is the confinement parameter. We note
that $H\ge 0$,  with $H=0$ allowed since $\Omega$ is the volume
available to the particles' \emph{centres}. In the case of a
narrow-channel, when $H<\epsilon$ particles cannot pass each other. We assume
that the volume 
fraction is small; since  $|\Omega| = \mathcal O(\epsilon^k)$
this implies that $N \epsilon^{d_e} \ll 1$.

\begin{figure}[htb]
\begin{center}
\input{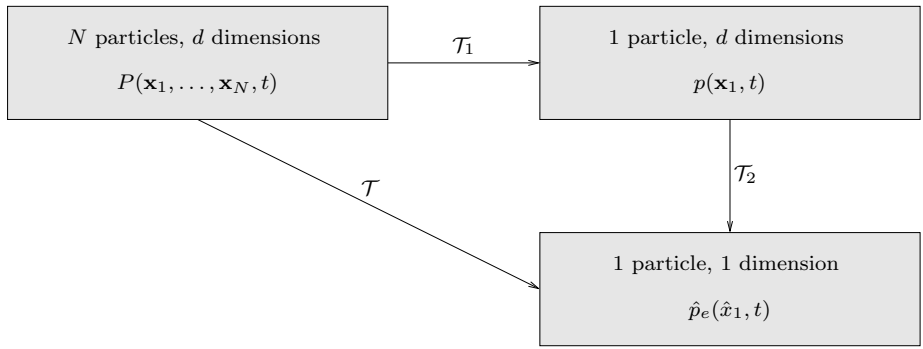}
\caption{Schematic of the problem solution steps for $d_e=1$ [narrow-channel cases (NC2) and (NC3)]. The goal is transformation $\mathcal T$, to obtain an effective one-dimensional equation along the channel for the marginal density of one particle. We achieve this with the combined steps $\mathcal T_1$ followed by $\mathcal T_2$.}
\label{fig:scheme}
\end{center}
\end{figure}
The high-dimensional diffusion problem \eqref{fp} will be reduced to
an effective $d_e$-dimensional transport model in two steps (see
\figref{fig:scheme}). First, as we did in \cite{Bruna:2012cg}, the
dimensions can be reduced from $dN$ to $d$ (individual to
population-level description) by looking at the marginal density
function of one particle (the first particle, say) given by  $p({\bf 
  x}_1,t) = \int P(\vec x, t) \, \ud {\bf x}_2 \cdots \ud {\bf x}_N$
(the particle choice is unimportant  since all the particles are
identical). Second, we will exploit the geometry of the domain
$\Omega$ to further reduce the dimensionality by $k$, the number of
confining dimensions.  
To this end, we will introduce the \emph{narrow-domain variables} and
obtain, from the $d$-dimensional density $p({\bf x},t)$ a reduced
\emph{effective} density $\hat p_e ({\bo x}_e,t)$, with ${\bo x}_e \in
\mathbb R^{d_e}$. For cases (NC2) and (NC3), the effective density
$\hat p_e$ will be a one-dimensional density $\hat p_e(x,t)$ along the
channel axis. For (PP), it will be an effective two-dimensional
density on the plane, $\hat p_e \equiv \hat p_e(x,y,t)$.   

For the sake of clarity we illustrate the derivation for the two-dimensional case (NC2) for both point and finite-size particles; the extension to the three-dimensional cases follows similarly and the respective models are only given in a summarised form.

\subsection{Point particles} \label{sec:points}

We begin by considering the case of point particles, for which the first reduction $\mathcal T_1$ in \figref{fig:scheme} from $N$ to one particle is straightforward. Since the particles are independent, $P(\vec x,t) = \prod_{i=1}^N p({\bf x}_i,t)$, and
\begin{subequations}
\label{npoint}
\begin{alignat}{3}
\label{npoint_eq}
\frac{\partial p}{\partial t}({\bf  x},t) &=  {\boldsymbol \nabla}_{
  {\bf x}} \cdot \left[ {\boldsymbol \nabla}_{{\bf  x}} \,  p -  
  {\bf f}({\bf  x}) \, p \right]& \qquad  & \textrm{in} &\qquad &\Omega,\\ 
\label{npoint_bc} 
0&= \left[ {\boldsymbol \nabla}_{{\bf  x}} \,  p -  {\bf  f}(
  {\bf x}) \, p \right] \cdot \boldsymbol { \hat {\bf n}} & &
\textrm{on} & & \partial \Omega, 
\end{alignat}
\end{subequations}
where $ \boldsymbol {\hat  {\bf n}}$ is the outward unit normal to
$\partial \Omega$. 
Thus we move to the second model reduction $\mathcal T_2$ which is applied to \eqref{npoint}. Using the definition of $\Omega$ \eqref{OmegaNC2}, we want to exploit the smallness of $H$. We introduce a change of variables to the \emph{narrow-domain variables}, which consist of rescaling by $\epsilon$ the variables corresponding to the confined dimension:
\begin{align}
\label{narrowvariables}
x = \hat x, \qquad  y= \epsilon \hat y.
\end{align}
Introducing $h$ such that $H = \epsilon h$, the domain $\Omega$ transforms into $\omega = \left[-\tfrac{1}{2}, \tfrac{1}{2}\right] \times \left[-\tfrac{h}{2}, \tfrac{h}{2}\right]$. 
In the rescaled domain, we define $\hat p(\hat {\bf x},t) = \epsilon p ({\bf  x}, t)$. 
(The factor of $\epsilon$ is introduced so that both $p$ and $\hat p$ integrate to one in their respective domains $\Omega$ and $\omega$.) Then \eqref{npoint} becomes
\begin{subequations}
\label{npointn}
\begin{align}
\label{npoint_eqn}
\epsilon ^2 \frac{\partial \hat p}{\partial t}(\hat {\bf  x}, t) = \epsilon ^2 \frac{\partial}{\partial \hat x} \left ( \frac{\partial \hat p}{\partial \hat x} - f_1(\hat x, \epsilon \hat y) \hat p \right) + \frac{\partial}{\partial \hat y} \left (   \frac{\partial \hat p}{\partial \hat y} - \epsilon f_2(\hat x, \epsilon \hat y) \hat p \right),
\end{align}
in $\omega$, with boundary conditions 
\begin{alignat}{3}
\label{np_bc1}
\frac{\partial \hat p}{\partial \hat x} &= f_1(\hat x, \epsilon \hat y) \hat p& \qquad & \text{on} & \qquad & \hat x = \pm \frac{1}{2},\\ 
\label{np_bc2}
\frac{\partial \hat p}{\partial \hat y} &=  \epsilon f_2(\hat x, \epsilon \hat y) \hat p& \qquad & \text{on} & \qquad & \hat y = \pm \frac{h}{2},
\end{alignat}
\end{subequations}
where $f_1$ and $f_2$ are respectively the horizontal and vertical components of the external force $\bf f$. 
Expanding $\hat p$ in powers of $\epsilon$, Taylor-expanding $f_1$ and $f_2$ around $(\hat x, 0)$, and solving \eqref{npoint_eqn} with the boundary condition \eqref{np_bc2} gives, at leading order, that $\hat p$ is independent of $\hat y$. Integrating \eqref{npoint_eqn}  over the channel's cross section and using \eqref{np_bc2} we find that, to $\mathcal O (\epsilon)$
\begin{align}
\label{ntotal}
\frac{\partial \hat p_e}{\partial t}(\hat {x}, t)  &=  \frac{\partial}{\partial \hat x} \left( \frac{\partial \hat p_e}{\partial \hat x} - f_1(\hat x, 0) \, \hat p_e  \right) \qquad \hat x \in [-1/2,1/2],
\end{align}
where $\hat p_e = \int_{-h/2}^{h/2} \hat p \, \ud \hat y$ is the
effective one-dimensional density along the channel. This equation is
complemented with no-flux boundary conditions at $\hat x = \pm
1/2$. Equation \eqref{ntotal} can be generalised to three-dimensional
geometries as 
\begin{align}
\label{ntotal3}
\frac{\partial \hat p_e}{\partial t}(\hat { \bf x}_e, t)  &=  \bo \nabla_{\hat {\bf x}_e} \cdot \left[   \bo \nabla_{\hat {\bf x}_e}  \hat p_e  - {\bf f}_e (\hat {\bf x}_e) \, \hat p_e  \right]  \qquad \hat {\bf x}_e \in \omega_e,
\end{align}
with no-flux boundary conditions on $\partial \omega_e$, where $\hat {
  \bf x}_e \in \omega_e$ are the coordinates in the
effective domain (i.e. the one-dimensional axis for (NC3) as in \eqref{ntotal},
or the two-dimensional plane for (PP)). The effective drift ${\bf f}_e$
is the projection of the full drift vector onto the effective domain
$\omega_e$. The initial condition is $\hat p_e(\hat {\bf x}_e, 0) =
\hat p_0 (\hat {\bf x}_e)$, where $\hat p_0 (\hat {\bf x}_e) =
\int_{\Omega^N} P_0 (\vec x) \delta(\hat {\bf x}_e - {\bf x}_{1,e})
\ud \vec x$.  

A common extension to \eqref{ntotal} is to suppose that the channel
has a non-constant cross section,  $h = h(x)$. The simplest model is
the Fick--Jacobs equation \citep{jacobs1967diffusion}, which in our
notation reads 
\begin{equation}
\label{fickjacobs}
\frac{\partial \hat p_e}{\partial t}(\hat x, t)  =  \frac{\partial}{\partial \hat x} \left[ h(\hat {x}) \frac{\partial }{\partial \hat x} \left( \frac{\hat p_e}{  h(\hat x)}  \right) - f_1(\hat x, 0) \, \hat p_e  \right],
\end{equation}
and is valid for $\epsilon
h'(x)$ small. Generalisations to this equation to account for the
channel curvature (a higher-order term) have been given in
\citep{Reguera:2001ev}.  
The key step in deriving \eqref{fickjacobs} is to assume that the full
two- or three-dimensional probability density $\hat p(\hat{\bf x},t)$
is at equilibrium in the transverse direction, that is, it is
assumed to factorise as 
$\hat p(\hat{\bf x},t)  \approx \hat p_e(\hat x,t) \rho(\hat {\bf x})$,
where $\rho(\hat {\bf x})$ is the local equilibrium distribution of
$\hat y$ (and $\hat z$, for $d=3$), conditional on a given $\hat x$
(the normalised Boltzmann--Gibbs probability density); see
\cite{Zwanzig:1992ta}. 

 In what follows, we keep $h$ constant since the inclusion of a
 variable channel width in the analysis for finite-size particles is
 not  straightforward.  

\subsection{Finite-size particles} \label{sec:finitesize}

We now describe the main result of this paper: the model of the
effective dynamics in a confined domain for the drift-diffusion of
finite-size particles. Using a similar technique to our previous work
\cite{Bruna:2012cg}, we are able to reduce the Fokker--Planck equation
\eqref{fp} for the joint probability density $P(\vec x, t)$ of $N$
interacting finite-size particles in a confined domain $\Omega$ to the
following effective equation for the marginal density $\hat p_e(\hat
{\bo x}_e, t)$: 
\begin{align}
\label{fp_reduced}
\frac{\partial \hat p_e}{\partial t}(\hat { \bo x}_e, t)  &=  \bo \nabla_{\hat {\bo x}_e} \cdot \left \{  \left[ 1+ (N-1) \epsilon^{d_e} \alpha_h \hat p_e \right] \bo \nabla_{\hat {\bo x}_e}  \hat p_e  - {\bo f}_e (\hat {\bo x}_e) \, \hat p_e  \right \},
\end{align}
for $\hat {\bo x}_e \in  \omega_e \subset \mathbb R^{d_e}$, where
$d_e$ are the effective dimensions of the reduced domain
$\omega_e$. The coefficient $\alpha_h$, which depends on the geometry
of the problem, determines how the excluded volume varies with the
confinement parameter $h$. This equation is complemented with no-flux
boundary conditions on $\partial \omega_e$ and initial data $\hat
p_e(\hat {\bo x}_e,0) = \hat p_0(\hat {\bo x}_e)$. Below we specify
the coefficient $\alpha_h$ for some specific cases.  
\begin{enumerate}
\item Two-dimensional channel of width $h$ (NC2), $d_e = 1$:
\begin{equation}
\label{functiong}
\alpha_h =  \frac{1}{h^2} \left[\pi h - \frac{4}{3}  + \Theta(1-h) \left( \frac{2}{3} (2 + h^2) \sqrt{1-h^2}  - 2h \arccos(h) \right) \right],
\end{equation}
where $\Theta(x)$ is the Heaviside step function,
\[ \Theta(x) = \left\{ \begin{array}{ll} 0 \qquad & x< 0,\\
1 & x \geq 0.
\end{array}
\right.
\]
\item Three-dimensional channel of cross-section $h \times h$ (NC3), $d_e = 1$:
\begin{align}
\label{functiongNC3}
\alpha_h = \frac{1}{h^4} \left[ \Theta(h-1) \left( \frac{4\pi}{3} h^2 - \pi h + \frac{8}{15} \right)  + \Theta(1-h) m(h) \right],
\end{align}
where
\begin{align}
m(h) = s(h) + \left \{ 
\begin{aligned}  
&\sigma_a(h)  &0  \le &h \le \tfrac{\sqrt 5 -1}{2},\\
&\sigma_b(h)  &\tfrac{\sqrt 5 -1}{2} < &h \le \tfrac{1}{\sqrt 2}, \\
&0 \qquad &\tfrac{1}{\sqrt 2} < &h , 
\end{aligned} \right.
\end{align}
with
\begin{align*}
s(h) &= \frac{8}{15} + \frac{2}{15}  \sqrt{1-h^2} (2 h^4 -9 h^2 -8)  - \frac{\pi}{3} h  \big(h^4 - 6 h^2+ 4h - 3\big)  - 2 h \arcsin(h),
\end{align*}
\begin{align*}
\sigma_a(h) &= \frac{2}{15} \sqrt{1-2 h^2} (h^4 +9 h^2 + 4)+ \frac{\pi}{12} h \left( 3 h^4 - 18 h^2 + 16 h - 9 \right) \\
& \phantom{=}+ \frac{1}{6} h^3 (h^2 -6 ) \text{arccot}\left (\frac{2 h \sqrt{1-2 h^2}}{1-3 h^2}\right) -\frac{4}{3} h^2 \text{arccot}\left (\frac{1- 2 h^2 - h^4 }{2 h^2 \sqrt{1-2 h^2}}\right)
\\
& \phantom{=}
-\frac{1}{2} h \text{arccot} \left ( \frac{2 h (1-2 h^2 )^{3/2}+2 h \sqrt{1-h^2} (3 h^2 - 1 )}{1-5 h^2+6 h^4+4 h^2 \sqrt{(1-2 h^2) (1-h^2)}}\right)
\\
&\phantom{=} + h \arcsin (h) -\frac{1}{3} h (h^4 -6 h^2 - 3 ) \arcsin \left (\frac{h}{\sqrt{1-h^2}}\right ), 
\end{align*}
\begin{align*}
\sigma_b(h) &=  \frac{2}{15} \sqrt{1-2 h^2} (h^4 +9 h^2 + 4)+ \frac{\pi}{12} h \left( 2 h^4 - 12 h^2 + 8 h - 3 \right) \\
&\phantom{=} + \frac{1}{3} h^3 (h^2 - 6 ) \text{arccot}\left( \frac{2 h \sqrt{1-2 h^2}}{1-3 h^2}\right) + \frac{4}{3} h^2 \arctan \left ( \frac{1-2h^2 - h^4 }{2 h^2 \sqrt{1-2 h^2}}\right) \\
&\phantom{=} + \frac{1}{2} h \, \text{arccot}\left ( \frac{4 h \sqrt{1-2 h^2} (3 h^2 - 1)}{1-10 h^2+17 h^4}\right).
\end{align*}
\item Three-dimensional parallel plates a distance $h$ apart (PP), $d_e = 2$:
\begin{align}
\label{functiongPP}
\alpha_h = \frac{\pi}{6 h^2 } \left[  \left(h^2(6-h^2) \right) \Theta(1-h)  +  (8h - 3)\Theta(h-1)\right].
\end{align}
\end{enumerate}
The coefficient $\alpha_h$ corresponding to cases (NC2), (NC3), and
(PP) is plotted in \figref{fig:alpha}.  
\def \scc {.9}
\def \scl {1.2}
\begin{figure}[ht]
\unitlength=1cm
\begin{center}
\psfragscanon
\psfrag{h}[][][\scl]{$h$}  \psfrag{a}[][][\scl]{$\alpha_h$ \ } 
\psfrag{data1}[][][\scc]{(NC2)} 
\psfrag{data2}[][][\scc]{(NC3)}  \psfrag{data3}[][][\scc]{(PP)} 
\includegraphics[width = .54\linewidth]{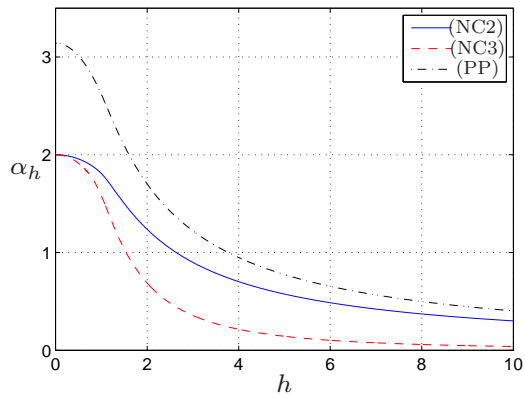}
\caption{Excluded-volume coefficient $\alpha_h$ as a function of the confinement parameter $h$ in three cases: two-dimensional channel (NC2) \eqref{functiong}, three-dimensional channel (NC3) \eqref{functiongNC3}, and parallel plates (PP) \eqref{functiongPP}.}\label{fig:alpha}
\end{center}
\end{figure}

Equation \eqref{fp_reduced} describes the probability density for
finding the first particle at position $\hat {\bf x}_e$ at time
$t$. Since the original system \eqref{fp} is invariant to permutations
of the particle labels, the marginal density function of any other
particle is the same. Thus the probability distribution function for
finding \emph{any} particle at position $\hat {\bo x}_e$ at time $t$
is simply $N\hat p_e$.  

\begin{figure}[htb]
\begin{center}
\resizebox{0.8\textwidth}{!}{\input{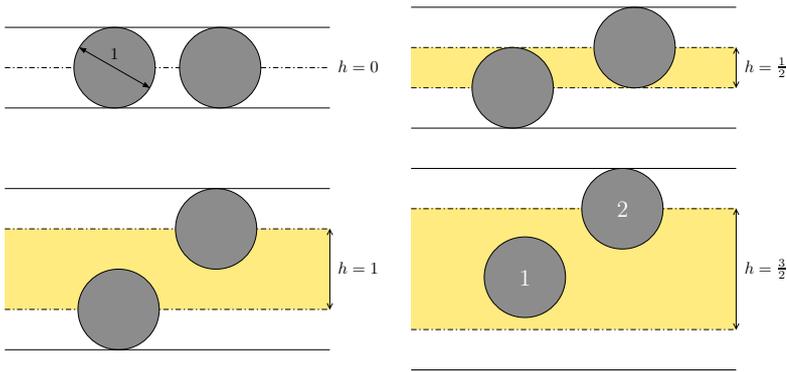}}
\caption{Sketch of the channel domain (shaded in yellow) for different values of $h$ with  particles of diameter one (note that here we are depicting the actual particles, not their excluded-area which has \emph{radius} one). Single-file channel for $0\le h <1$ (with $h=0$ being the extreme case in which particles can only move in the axial direction). When $h=1$ particles can \emph{just} pass each other, and for $h>1$ (bottom row) particles can more easily change order. }
\label{fig:zones}
\end{center}
\end{figure}

 In \figref{fig:zones} we sketch the narrow-channel domain (rescaled by
$\epsilon$) for various heights $h$. The physical domain (of width
$h+1$ in the narrow-domain variables) is delimited by the solid black
lines, while the configuration domain (of width $h$) corresponds to
the yellow shaded region delimited by dot-dash lines.   

The nonlinear diffusion term in \eqref{fp_reduced} is proportional to
the {\em effective} excluded volume created by the remaining $(N-1)$
particles as well as the domain walls after the dimensional
reduction. 
For example, in the (NC2) case $\epsilon \alpha_h$
 is the effective one-dimensional excluded \emph{interval},
corresponding to the excluded area divided by the height of the cross
section available to a particle centre (see \figref{fig:zones}). 
When $h=0$, a particle of diameter
$\epsilon$ excludes an interval of $2\epsilon$ (this explains why
  $\alpha_h = 2$ for $h=0$; see \figref{fig:alpha}). As the channel
width increases, the value $\epsilon \alpha_h$ decreases since the
whole width of the channel is not always excluded by a given particle.
As $h$ gets large $\epsilon \alpha_h$ gives the ratio of the
area excluded by the particle, $\pi \epsilon^2$, to the cross section
height $\epsilon h$, so that  $\alpha_h \sim \pi/h$ as $h\to \infty$
[see \eqref{functiong}].  

While $\alpha_h$ gives the effective excluded volume after dimensional
reduction, the  actual excluded volume is proportional to $h \alpha_h$.  This is 
plotted in \figref{fig:alphah} [along with the corresponding
expressions for (NC3) and (PP)].

\def \scc {0.9}
\def \scl {1.0}
\begin{figure}[ht]
\unitlength=1cm
\begin{center}
\psfragscanon
\psfrag{h}[][][\scl]{$h$}  \psfrag{ah}[][][\scl]{$\alpha_h A \quad$} 
\psfrag{data1}[][][\scc]{(NC2)} 
\psfrag{data2}[][][\scc]{(NC3)}  \psfrag{data3}[][][\scc]{(PP)} 
\includegraphics[width = .54\linewidth]{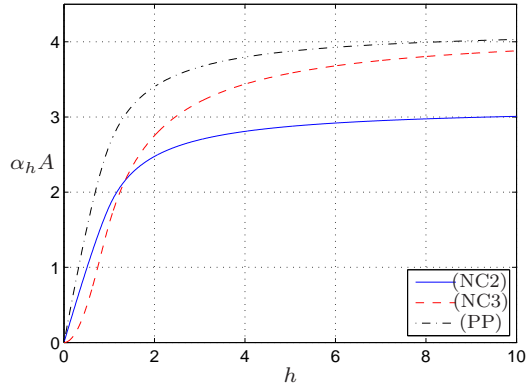}
\caption{Excluded volume  $\alpha_h A$ as a function of the confinement parameter $h$ in three cases: two-dimensional channel (NC2) ($A=h$), three-dimensional channel  (NC3) ($A=h^2$), and parallel plates (PP) ($A=h$).}\label{fig:alphah}
\end{center}
\end{figure}

It is clear from  \figref{fig:zones}  that the excluded volume due to 
a particle varies depending on 
its position in the channel's cross-section: in (NC2), while a
particle excludes an area of $\pi$ when it is far from the channel
walls, it only excludes half of this area when in contact with the
channel walls (less if $h<1$). This effect, known as an {\em entropic
  effect}, implies 
that the average excluded area over possible locations across the
channel width decreases as the channel narrows ($h\to 0$). As the
channel width $h$ grows, the boundary effects in which the excluded
area is reduced contribute less and less to the average value,
implying that the average excluded area tends to the constant value
$\pi$ as $h \to \infty$, which corresponds to the ``bulk'' excluded
area. This is confirmed in \figref{fig:alphah}. Similarly, $\alpha_h$
tends to $4\pi/3$ for the three dimensional cases as this is the
rescaled excluded volume (the volume of the unit sphere).  As $h \rightarrow
0$ the average excluded area $h \alpha_h \rightarrow 0$ since in the
extreme confinement cases almost all of the actual excluded area lies 
\emph{outside} the domain available to a particle's centre. 

\subsubsection{Effective equation for the volume concentration} \label{sec:eq_for_conc}

In our derivation of \eqref{fp_reduced} we do not require $N$ to be
large: in fact, equation \eqref{fp_reduced}  is valid for any $N$ (as
long as the volume fraction is small), so that one could set $N=1$ or
2 if required. This equation gives the \emph{probability} of finding a
particle at a given position at a given time. However, for large $N$
such that $N-1 \approx N$  we can introduce the volume concentration
$\hat c_e = \phi \hat p_e$, where $\phi$ is the total volume fraction
of particles,
and rewrite   equation \eqref{fp_reduced} as an  equation
for the concentration of particles in the system:\footnote{Note the factors $(1+h)$ in $\phi$: this is because $\phi$ is the
total volume of particles divided by the  actual volume of the
channel, not the volume available to a particle's centre.} 
\begin{align}
\label{reduced_conc}
\frac{\partial \hat c_e}{\partial t}(\hat { \bo x}_e, t)  &=  \bo \nabla_{\hat {\bo x}_e} \cdot \left [ ( 1+  g_h \hat c_e ) \bo \nabla_{\hat {\bo x}_e}  \hat c_e  - {\bo f}_e (\hat {\bo x}_e) \, \hat c_e  \right ],
\end{align}
with
\begin{subequations}
\label{extremes}
\begin{alignat}{3}
\label{extreme1}
&\text{(NC2)}:& \qquad g_h &= \frac{4}{\pi} (h+1) \alpha_h,& \qquad \phi &= \frac{N \pi \epsilon}{4 (h+1)},\\
\label{extreme2}
&\text{(NC3)}:& \qquad g_{h} &= \frac{6}{\pi} (h+1)^2\alpha_{h},& \qquad \phi &= \frac{N \pi \epsilon}{6 (h+1)^2},\\
\label{extreme3}
&\text{(PP)}:& \qquad g_{h} &= \frac{6}{\pi} (h+1)\alpha_h,& \qquad \phi &= \frac{N \pi \epsilon^2}{6 (h+1)}.
\end{alignat}
\end{subequations}
These expressions for $g_h$ are plotted in \figref{fig:extreme}. We note all three have a finite value as both $h\rightarrow 0$ and $h \rightarrow
\infty$ and, most 
importantly, that they have a relative maximum at $h = h^*$, where $h^*$ is slightly greater than one. 
In particular, $h^* = 1.47$, $h^* = 1.28$, and $h^* = 1.2$ for (NC2),
(NC3), and (PP) respectively. 
\def \scc {0.9}
\def \scl {1.0}
\begin{figure}[bh]
\unitlength=1cm
\begin{center}
\psfragscanon
\psfrag{h}[][][\scl]{$h$}  \psfrag{ex}[][][\scl]{$g_h$} 
\psfrag{data1}[][][\scc]{(NC2)} 
\psfrag{data2}[][][\scc]{(NC3)}  \psfrag{data3}[][][\scc]{(PP)} 
\includegraphics[width = .56\linewidth]{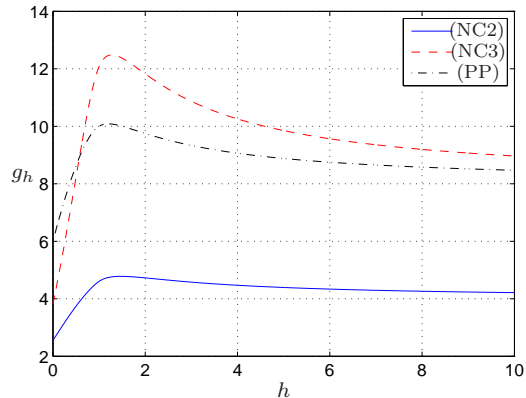}
\caption{Coefficient $g_h$ as a function of the confinement parameter $h$ in three cases: two-dimensional channel (NC2), three-dimensional channel (NC3), and parallel plates (PP).}\label{fig:extreme}
\end{center}
\end{figure}

The presence of this relative maximum at a fixed volume fraction is
interesting, as it implies an optimal ratio between the particles' size
and the confinement dimension at which excluded-volume effects, and
thus the effective transport, are maximised. In terms of the
physical domains, it corresponds to a narrow domain around $2.2$ to $2.5$
times wider than the particles' diameter, so that two particles can
just diffuse side by side in the channel.  
\begin{figure}[thb]
\begin{center}
\input{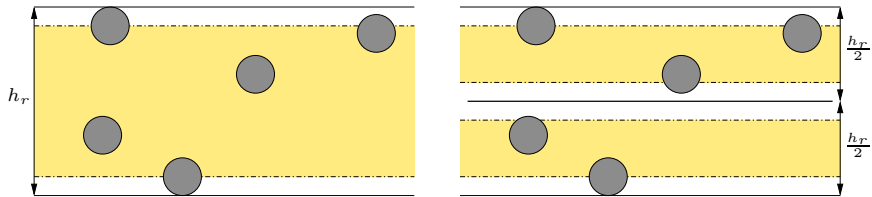}
\caption{Sketch of channel of  width $5 \epsilon$  with and
  without an intermediate wall that creates two lanes. The domain
  available to the particles centres' in each case is shaded in
  yellow. Left: one narrow channel with $h = 4$. Right: two narrow
  channels each with $h = 1.5$ (roughly equal to $h^*$ for maximal
  exclusion effects).} 
\label{fig:lanes}
\end{center}
\end{figure}

Thus the theory predicts that diffusive transport in a (NC2) channel of
width $5 \epsilon$ ($h=4$) may be increased by dividing the channel
into two sub-channels of width $2.5 \epsilon$ ($h = 1.5$), as shown in
\figref{fig:lanes}. This operation gives an
increase in $g_h$  of 7\%. The increase is more dramatic in the
three-dimensional case: if a (NC3) square channel has an 
original  width and depth of  $4.6 \epsilon$ (so that $h = 3.6$), then
subdividing it into four identical channels of  width $2.3
\epsilon$ (so that each has $h = 1.3$) gives a relative
increase in  $g_h$ is of 19\%. 

\section{Limiting cases: from single-file diffusion to unconfined
  diffusion}  \label{sec:limits} 

We have briefly discussed the limiting behaviour of $\alpha_h$ as $h
\rightarrow 0$ and $h \rightarrow \infty$ above. Here we examine these
limits in  equation \eqref{fp_reduced}, and check that they agree with
existing  results.
For  a channel of width  $h$ (NC2 or NC3) equation \eqref{fp_reduced}
interpolates between two limiting cases: a single-file channel ($h
\rightarrow 0$) and an unconfined two- (NC2) or three-dimensional (NC3)
domain ($h \to 
\infty$). For the (PP) case, the limit $h\to 0$ gives an (unconfined)
two-dimensional diffusion, while the  limit $h \to \infty$ gives 
three-dimensional unconfined diffusion.  
Finally, the extension of the square cross section (NC3) to a
rectangular cross section $h \times m$ can be used to interpolate
between (NC2) (as $m \rightarrow 0$) and (PP) (as $m \to \infty$). 

In this section we will examine these limits by comparing the limiting
behaviour of our model \eqref{fp_reduced} with the limiting problem of
diffusion of hard spheres in $\mathbb R^d$ for $d = 1 ,2, 3$. This
problem has been studied extensively, especially in the
one-dimensional case, which is known as \emph{single-file} diffusion
\citep{Lizana:2009ic}. For the cases $d=2,3$ we will use the results
from our previous work of unconfined diffusion of hard spheres
\citep{Bruna:2012cg}.  

\subsection{Limit to an unconfined domain: $h \to \infty$} \label{sec:hinfty}
 
As  $h$ increases we
expect the boundary effects contained in $\alpha_h$ to vanish and to
recover the ``bulk'' or unconfined equation found in \cite{Bruna:2012cg}: 
\begin{align}
\label{fpN_bulk} 
\frac{\partial p}{\partial t}({\bf x}, t) &= \nabla_{{\bf x}} \cdot \left \{ [1 +  (N-1) \alpha \epsilon ^d p ] \nabla_{{\bf x}} p - {\bf f}({\bf x}) p \right \} \quad \text{in} \quad \Omega,
\end{align}
where $\Omega \subset \mathbb R^d$ as given in \eqref{Omeganarrow}, $\alpha= \pi$ for $d=2$ and $\alpha = 4 \pi /3$ for $d=3$ (or $\alpha = 2(d-1) \pi /d$ for $d=2,3$). It is important to note that this equation is only valid for $H = \mathcal O(1)$ (that is, when $\Omega$ has volume order one). 

In order to take the limit $h\to \infty$ in our model \eqref{fp_reduced}, it is convenient to use the original density $\hat p$ in $\mathbb R^d$ rather than the effective density $\hat p_e$. In other words, we consider the following equation for $\hat p = \hat p_e /A$ (where  $A$ is the cross-sectional area):
\begin{align}
\label{fp_limitinf}
\frac{\partial \hat p}{\partial t}(\hat { \bo x}_e, t)  &=  \bo \nabla_{\hat {\bo x}_e} \cdot \left \{  \left[ 1+ (N-1)  \alpha_h A  \epsilon^{d_e} \hat p \right] \bo \nabla_{\hat {\bo x}_e}  \hat p  - {\bo f}_e (\hat {\bo x}_e) \, \hat p  \right \},
\end{align}
where $\alpha_h A$ is the excluded-volume coefficient shown in
\figref{fig:alphah}. As  can be seen in the figure [or in the formulas
for $\alpha_h$  \eqref{functiong}--\eqref{functiongPP}], the limit of
$\alpha_h A$ as $h \to \infty$ is $\pi$ for the two-dimensional
channel (NC2) and $4 \pi /3$ for the three-dimensional cases (NC3) and
(PP).  
Therefore, the limiting behaviour of \eqref{fp_limitinf} as $h \to \infty$ corresponds to replacing $\alpha_h A $ by $\alpha$, where the latter is given in the bulk equation \eqref{fpN_bulk}. 
The last step to show that \eqref{fpN_bulk} is indeed the limiting
model of \eqref{fp_limitinf} is to integrate \eqref{fpN_bulk} over the
cross section to reduce it to a $d_e-$dimensional equation as
\eqref{fp_limitinf}. In other words, the limit $h\to \infty$ of the
confined-domain model \eqref{fp_limitinf} should coincide with the
limit $H \to 0$ of the bulk equation \eqref{fpN_bulk}.   
Rescaling the confined dimensions by $\epsilon$ [cf. \eqref{narrowvariables}] and integrating \eqref{fpN_bulk} over the cross-section, it is straightforward to arrive at the following equation for $\hat p   = \epsilon^k p$:
\begin{align}
\label{bulk_limit}
\frac{\partial \hat p}{\partial t}(\hat { \bo x}_e, t)  &=  \bo \nabla_{\hat {\bo x}_e} \cdot \left \{  \left[ 1+ (N-1)  \alpha  \epsilon^{d_e} \hat p \right] \bo \nabla_{\hat {\bo x}_e}  \hat p  - {\bo f}_e (\hat {\bo x}_e) \, \hat p  \right \},
\end{align}
where we have used no-flux boundary conditions on the cross-section boundaries.

\subsection{Limit to single-file diffusion: $h \to 0$} \label{sec:hto0}

We now consider the limiting case $h\to 0$ in the narrow channel cases
(NC2) and (NC3). From the plot of $\alpha_h$ in \figref{fig:alpha} we
have that $\lim_{h\to 0} \alpha_h =2$ in both cases. This is confirmed
by taking the limit $h\to 0$  in \eqref{functiong} for (NC2) or
\eqref{functiongNC3} for (NC3). Thus the
single-file limit of  \eqref{fp_reduced} is
\begin{equation}
\label{p_singlefilelimit}
\frac{\partial \hat p_e}{\partial t}(\hat x, t)  =  \frac{\partial}{\partial \hat x} \! \left[ \big[1 + 2(N-1)\epsilon \hat p_e \big]  \frac{\partial \hat p_e}{\partial \hat x} - f_1(\hat  x)  \hat p_e \right].
\end{equation}
We see that the effective diffusion coefficient for $N$ large (such that $N-1 \approx N$) is $D^c(c) = (1 + 2 c)$ with $c = N \epsilon \hat p_e$ being the particle concentration, which is consistent with that derived by \cite{Ackerson:1982ti} for a uniform particle concentration $c = N\epsilon$.

We now compare \eqref{p_singlefilelimit} 
with the classic  one-dimensional model of diffusing  hard rods. It is
well known 
that the one-dimensional diffusion of finite-size particles can be
mapped onto a point--particle problem (cf.
\citealp{Lizana:2009ic}). Using this trick, a fast diffusion equation
for the evolution for the marginal density of $N$ rods of length
$\epsilon$ under no external force ($f_1 \equiv 0$) and in the
thermodynamic limit ($N\to \infty$, $L\to \infty$, $N/L \to \phi$
finite) is found in \cite{Rost:1984ts} (in French, see
\citealp{Bodnar:2005kv} for an explanation of Rost method in English), namely 
\begin{equation}
\label{fastdiffusion}
\frac{\partial \rho}{\partial t}(\hat x, t)  =  \frac{\partial}{\partial \hat x} \! \left( \frac{1}{(1-\epsilon \rho)^2} \frac{\partial \rho}{\partial \hat x}  \right),
\end{equation}
where $\rho= N \hat p_e$ is the number density. Expanding the equation
above in $\epsilon$  we obtain, to $\mathcal O(\epsilon)$, 
\begin{equation}
\label{fastdiffusion2}
\frac{\partial \hat p_e}{\partial t}(\hat x, t)  =
\frac{\partial}{\partial \hat x} \! \left( (1 + 2 N \epsilon\hat p_e )
  \frac{\partial \hat p_e }{\partial \hat x}  \right), 
\end{equation}
which is in agreement with the large $N$ limit of 
\eqref{p_singlefilelimit}. An
alternative derivation of \eqref{p_singlefilelimit} using matched
asymptotics on the original problem (without elimination of the
hard-core parts) can be found in \cite{Bruna:2012ub}. It  differs from
that in \cite{Rost:1984ts} in that it is valid for any $N$ and allows
an external force field $\bf f$. 

\subsection{Other limits} \label{sec:otherlimits}

From \eqref{functiongPP} we see that $\lim_{h \to 0} \alpha_h = \pi$
(see also \figref{fig:alpha}), from which it is straightforward to
show that the limit $h\to 0$ of (PP)
corresponds to an unconfined two-dimensional diffusion. 

A final limit to consider concerns the generalisation of the
three-dimensional channel (NC3) to a  rectangular cross section $h \times
m$. We have already seen above
that this tends to  the single-file diffusion model for $h= m \to 0 $,
and to an unconfined three-dimensional diffusion for $h, m \to
\infty$. Now, keeping $h$ fixed, the extra parameter $m$ will allow us
to interpolate between a two-dimensional narrow channel (NC2) of width
$h$ as $m \to 0$ and two parallel plates a distance $h$ apart (PP) as $m\to
\infty$. For $m \ge 1$ it can be shown that the rectangular counterpart of
\eqref{functiongNC3} reads: 
\begin{equation}
\alpha_{hm} = \frac{1}{h^2 m ^2}\left\{ 
\frac{8}{15} + \Theta(h-1) \left[  \frac{4 \pi}{3} hm - \frac{\pi}{2}(h+m)  \right] + \Theta(1-h)  s(h, m) \right\} ,
\end{equation}
with 
\[
s(h, m) = \pi m h^2 \left ( 1 - \frac{h^2}{6} \right) - h \arcsin h  + \frac{ \sqrt{1-h^2}}{15} ( 2 h^4 - 9 h^2 - 8).
\]
Now, in order to compare between the one-dimensional (NC3) model and the two-dimensional (PP) model, the relevant quantity is $m \alpha_{hm}$ (so that the one-dimensional effective density $\hat p_e$ is mapped onto a two-dimensional plate of width $m$). We find that 
\begin{align*}
\lim_{m\to \infty} m \alpha_{hm}^\text{(NC3)} = \alpha_h ^\text{(PP)},
\end{align*}
as expected.

\section{Model analysis}

\subsection{Numerical analysis of time dependent solutions}\label{sec:time-dep}

In this section we compare solutions of our effective nonlinear
diffusion equation for the (NC2) case with direct stochastic
simulations, averaging over $2 \cdot 10^4$
stochastic realisations of the corresponding individual-based model
\eqref{sde}.  
For the PDE, we use the method of lines with a standard second-order
finite-difference discretisation of the spatial derivatives. For the
coupled system of SDEs, we perform Monte Carlo (MC) simulations of the
discretised version of \eqref{sde} using the Euler--Maruyama method,  
\begin{equation}
\label{sded}
{\bf X}_i( t+ \Delta t) = {\bf X}_i (t) + {\bf f}({\bf X}_i(t)) \Delta t + \sqrt{2 \Delta t} \, {\bo \xi}_i,
\end{equation} 
where ${\bo \xi}_i$ is a 2-vector whose entries are independent
normally distributed random variables with zero mean and unit
variance. The reflective boundary conditions on $\partial \Omega$
implemented as in \cite{Erban:2007we}, namely, the distance that a
particle has travelled outside the domain is reflected back into the
domain. Care must be taken for very narrow channels to account for the
possibility that a particle has travelled outside the domain by more
than a width $h$. The particle-particle overlaps are implemented
similarly: the difference $\epsilon - \| {\bf X}_i(t+ \Delta t) - {\bf
  X}_j(t + \Delta t) \|$ corresponds to the distance that particles
have penetrated each other illegally. Then we suppose that each
particle has travelled the same illegal distance, and we separate them
accordingly along the line joining the two particles' centres. This
approach works well for low volume fractions, but may run into
difficulty at high volume fractions when the separated particles may
suffer further overlaps. In that case the algorithm in
\cite{Scala:2007fd}, an event-driven Brownian dynamics, becomes more
suitable.

\begin{figure}[h!]
\def \scp {0.8}
\unitlength=1cm
\begin{center}
\psfragscanon
\psfrag{a}[][][\scl]{$(a)$}  \psfrag{b}[][][\scl]{$(b)$}  \psfrag{in}[l][][\scl]{$t=0$}  \psfrag{fi}[l][][\scl]{$t_{\! f}$}  \psfrag{data1}[][][\scp]{\eqref{fp_reduced}} \psfrag{data2}[][][\scp]{\eqref{p_singlefilelimit}} \psfrag{data3}[][][\scp]{\eqref{bulk_limit}} \psfrag{data4}[][][\scp]{\eqref{ntotal}} \psfrag{data5}[][][\scp]{\eqref{sde}}
\psfrag{x}[][][\scl]{$\hat x$}  \psfrag{p}[r][][\scl][-90]{$\hat p_e$} 
\includegraphics[width = .48\linewidth]{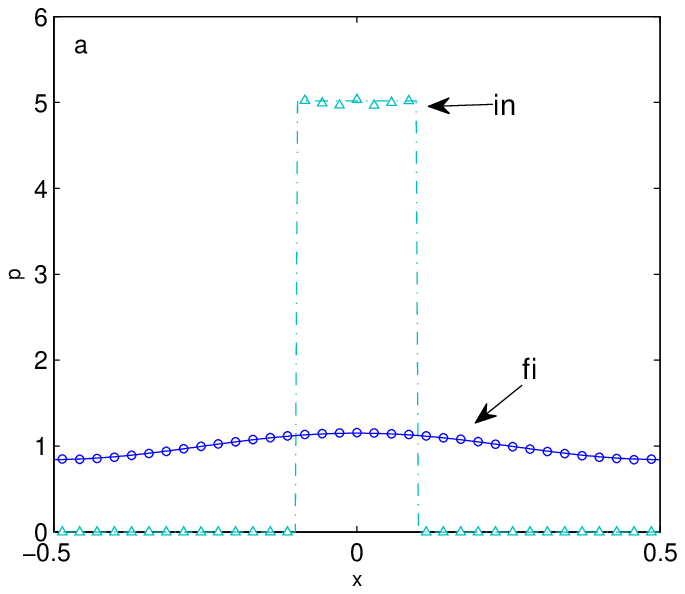} \quad
\includegraphics[width = .48\linewidth]{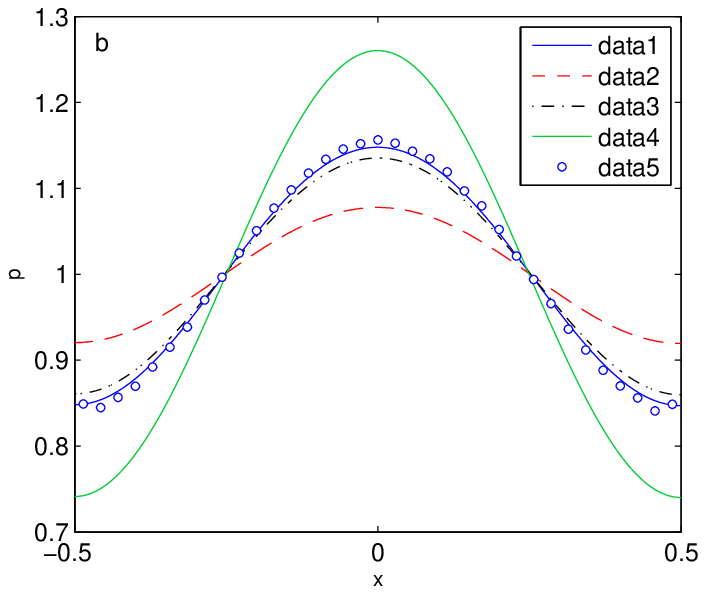}
\caption{One-dimensional density $\hat p_e(\hat x, t)$ in a channel of
  width $\epsilon h$ with no-flux boundary conditions at $\hat x = \pm
  0.5$: solution of the continuum model \eqref{fp_reduced} for
  finite-size particles versus individual-based
  model simulations of \eqref{sde} and limiting continuum models. (\textbf{a}) Initial ($t=0$) and final ($t_{\! f}=0.05$) densities $\hat p_e$ (lines) and histograms (data points). (\textbf{b}) Final density and histogram, together with three limiting cases: point particles or standard linear diffusion \eqref{ntotal},  unconfined limit  \eqref{bulk_limit} and single-file limit \eqref{p_singlefilelimit}. Parameters are $h=3$, $\epsilon = 0.01$, and $N=30$, $\Delta t = 10^{-5}$}
\label{fig:tdexample1}
\end{center}
\end{figure}
\begin{figure}[ht]
\def \scp {0.8}
\unitlength=1cm
\begin{center}
\psfragscanon
\psfrag{a}[][][\scl]{$(a)$}  \psfrag{b}[][][\scl]{$(b)$} \psfrag{data1}[][][\scp]{\eqref{fp_reduced}} \psfrag{data2}[][][\scp]{\eqref{p_singlefilelimit}} \psfrag{data3}[][][\scp]{\eqref{bulk_limit}} \psfrag{data4}[][][\scp]{\eqref{ntotal}} \psfrag{data5}[][][\scp]{\eqref{sde}}  \psfrag{in}[r][][\scl]{$t=0$}  \psfrag{fi}[][][\scl]{$t_{\! f}$}
\psfrag{x}[][][\scl]{$\hat x$}  \psfrag{p}[r][][\scl][-90]{$\hat p_e$} 
\includegraphics[width = .48\linewidth]{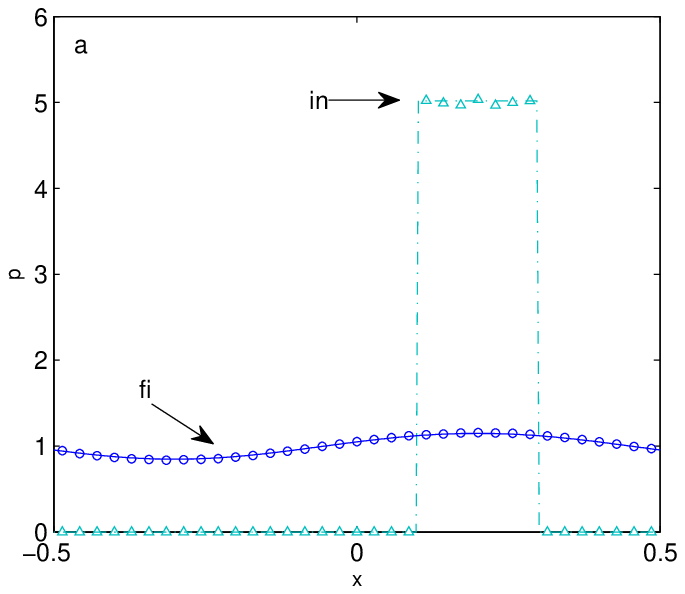} \quad
\includegraphics[width = .48\linewidth]{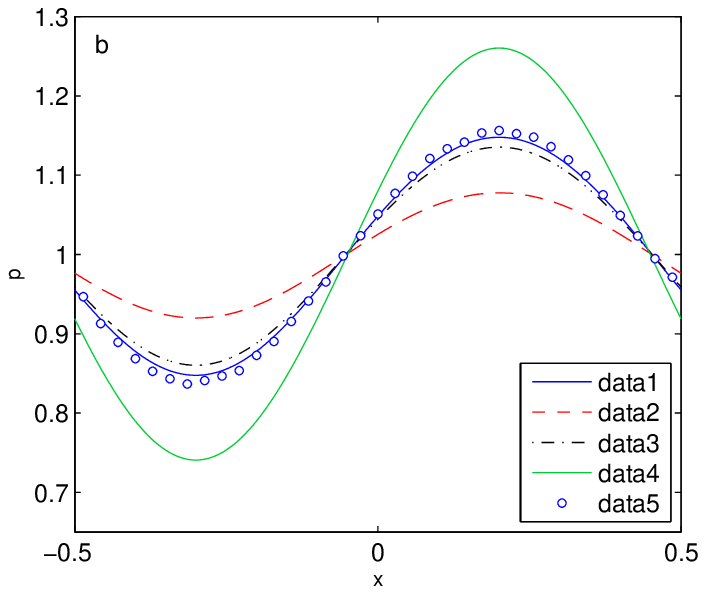}
\caption{One-dimensional density $\hat p_e(\hat x, t)$ in a channel of width $\epsilon h$ with periodic boundary conditions at $\hat x = \pm 0.5$: solution of the continuum model \eqref{fp_reduced} for
  finite-size particles versus individual-based
  model simulations of \eqref{sde} and limiting  models \eqref{ntotal},  \eqref{bulk_limit} and \eqref{p_singlefilelimit}. Other details of the plots are given in the caption of Fig. \ref{fig:tdexample1}. (Color figure online)}
\label{fig:tdexample2}
\end{center}
\end{figure}

Figures \ref{fig:tdexample1} and \ref{fig:tdexample2}  show the
numerical results at $t=0.05$ for $h=3$, $\epsilon=0.01$, $N=30$,
${\bf f} = 0$ with no-flux and periodic boundary conditions at the
channel ends, respectively. At initial time, the particles are
uniformly distributed in a segment of length 0.2 [Figures
\ref{fig:tdexample1}(a) and \ref{fig:tdexample2}(a) in cyan triangles
and dash line]. The data points show the one-dimensional histogram
obtained by averaging the MC results over the channel's cross section.  
To test the importance of the excluded-volume interactions and the
confinement, we also compare with the corresponding solutions with
point particles (equivalent to standard linear diffusion) and the limiting cases as $h\to  \infty$ and $h\to 0$
of Subsections \ref{sec:hinfty} and \ref{sec:hto0} respectively. 

In both cases we see very good agreement between the stochastic average and the solution of the narrow-channel model $\hat p_e$,  whilst there are noticeable differences between the three limiting models, namely the point particles, single-file and unconfined limits. In order to quantify the error committed by the limiting models, we note that, for the chosen parameters, the volume fraction is $\phi \approx 6\%$ and the nonlinear coefficient in \eqref{extreme1} is $g_h \approx 4.6$. The corresponding nonlinear coefficient in the limiting cases is 0 for the point particles limit, 2 for the single-file limit and 4 for the unconfined limit. The difference between the narrow channel coefficient and these limiting values are consistent with the differences observed in the numerical solutions: while for $h=3$ the unconfined limit compares reasonably well with the stochastic simulations, the single-file and the point particles limits show more important differences. However, we note that our model gives the best agreement with the MC simulations results.

The results of Figures \ref{fig:tdexample1} and \ref{fig:tdexample2} suggest that, depending on the value of $h$, either the single-file or the unconfined limits will be more appropriate, while our narrow-channel model captures the whole range of $h$ from 0 to $\infty$. To investigate this further, in \figref{fig:fge_nl1} we examine how the narrow-channel model compares with three limiting cases [point particles \eqref{ntotal}, unconfined \eqref{bulk_limit}, and single-file \eqref{p_singlefilelimit}] for various values of the channel width $h$ while keeping the volume fraction $\phi$ fixed.\footnote{We keep the volume fraction $\phi = N \pi \epsilon/4(h+1)$ fixed by varying $\epsilon$ as $h$ changes.}  It may seem counterintuitive that the limiting solutions appear to change with $h$ more than the narrow channel solution (solid blue line) does. However, we note that, although the three limiting models are independent of $h$, the unconfined model and the single-file model  solutions (shown in dot-dash black and dash red lines respectively in \figref{fig:fge_nl1}) do vary as $h$ increases because the value of $\epsilon$ is being changed in order to keep $\phi$ fixed. The narrow channel solution  also changes (albeit less markedly) with $h$ as predicted by the coefficient $g_h$ (see \figref{fig:extreme}).
\def \scc {0.7}
\def \scl {1.0}
\begin{figure}[ht!]
\def \scc {0.7}
\def \scl {1.0}
\def \scp {0.8}
\unitlength=1cm
\begin{center}
\psfragscanon
\psfrag{x}[][][\scl]{$\hat x$}  \psfrag{p}[][][\scl][-90]{$\hat p_e \ $} 
\psfrag{h = 0.5}[][][\scl]{$h = 0.5$}
\psfrag{h = 1}[][][\scl]{$h = 1$}
\psfrag{h = 1.5}[][][\scl]{$h = 1.5$}
\psfrag{h = 2}[][][\scl]{$h = 2$}
\psfrag{h = 3}[][][\scl]{$h = 3$}
\psfrag{h = 5}[][][\scl]{$h = 5$}
\psfrag{data4}[][][\scp]{\eqref{fp_reduced}} \psfrag{data1}[][][\scp]{\eqref{p_singlefilelimit}} \psfrag{data2}[][][\scp]{\eqref{bulk_limit}} \psfrag{data3}[][][\scp]{\eqref{ntotal}} 
\includegraphics[width = .38\linewidth]{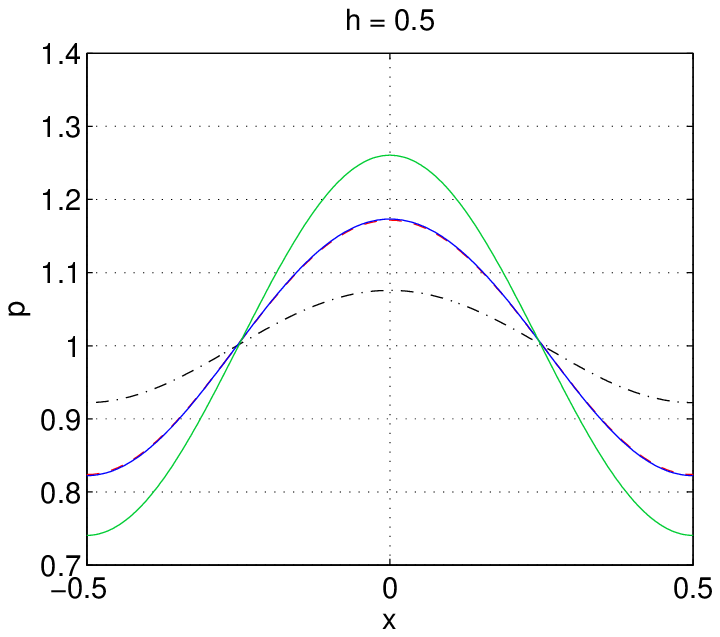} \quad
\includegraphics[width = .38\linewidth]{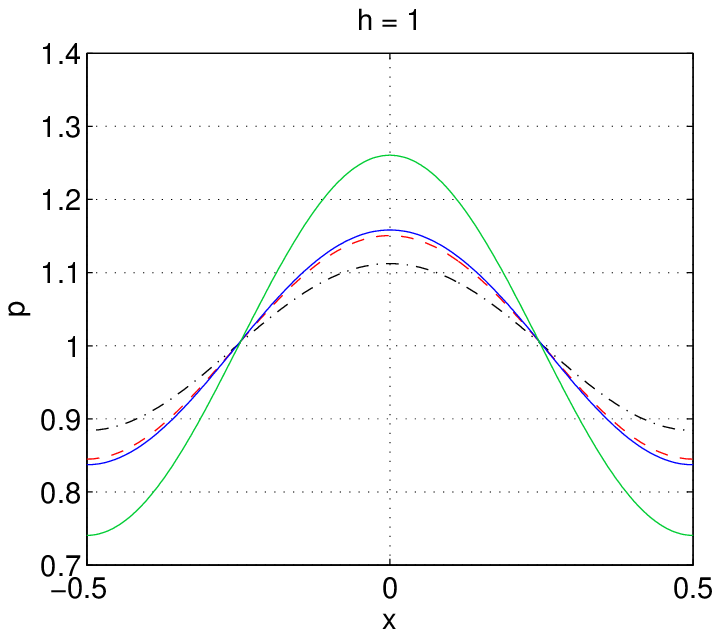} \\
\includegraphics[width = .38\linewidth]{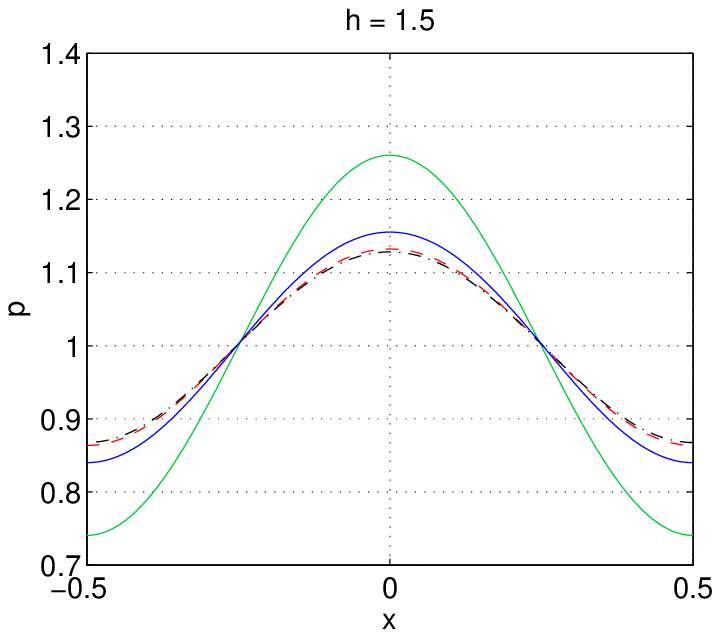} \quad
\includegraphics[width = .38\linewidth]{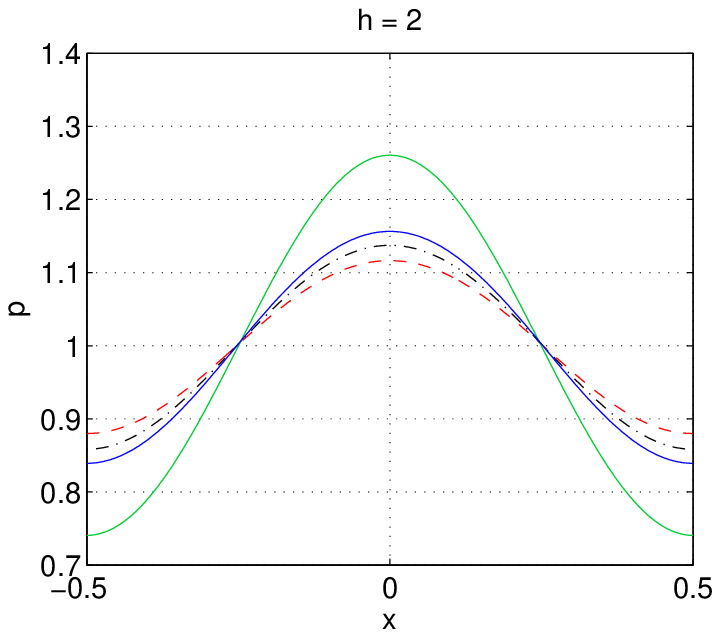} \\
\includegraphics[width = .38\linewidth]{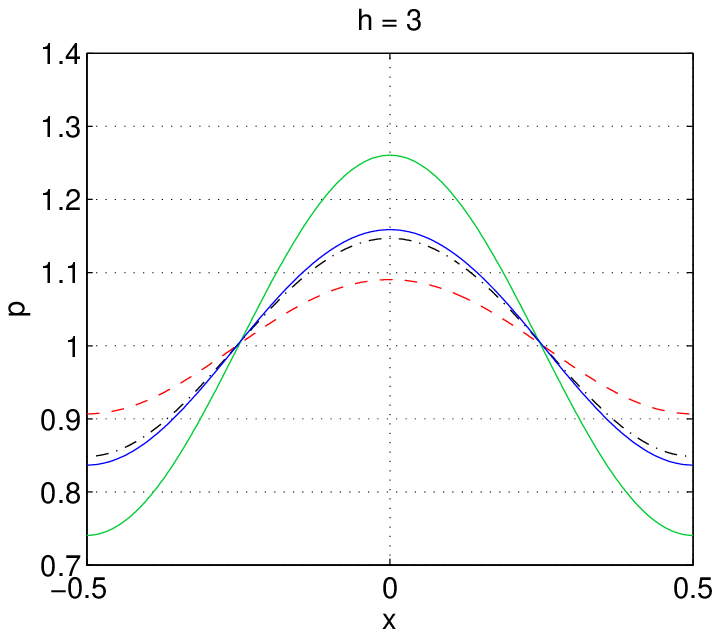} \quad
\includegraphics[width = .38\linewidth]{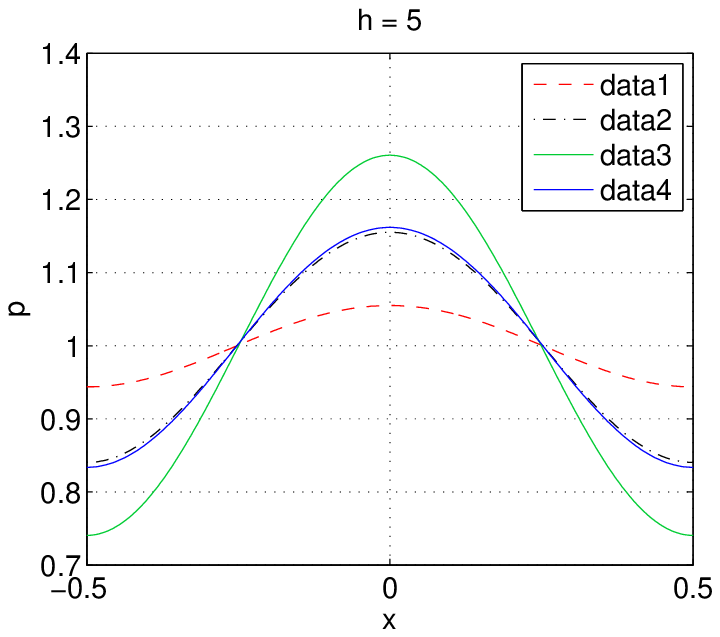} 
\caption{One-dimensional density $\hat p_e(\hat x, t)$ in a two-dimensional channel of width $\epsilon h$ with fixed volume fraction $\phi =0.05$ at time $t=0.05$. No-flux boundary conditions at $\hat x = \pm 0.5$ and uniformly initial conditions in $|\hat x| \le 0.1$. Solution $\hat p_e$ of the narrow-channel equation \eqref{fp_reduced} (solid blue), versus the single-file limit \eqref{p_singlefilelimit} (dash red line), the unconfined limit \eqref{bulk_limit}  (dot-dash black line, curve shown is $h \hat p$), and the point particles limit \eqref{ntotal} (solid green line). (Color figure online)}
\label{fig:fge_nl1}
\end{center}
\end{figure}

As expected, for $h=0.5$ the narrow-channel solution agrees very well with the single-file limit, since in this case the cross-sectional space is not enough to let particles pass each other (see top right plot in \figref{fig:zones}). In contrast, the unconfined case solution is far apart from the previous two, since the approximation that boundary effects are negligible is poor for $h=0.5$. As we increase $h$, we can observe how the single-file solution (in dash red)  moves apart from the narrow-channel solution (in solid blue), while the unconfined case solution (in dot-dash black) becomes closer to the latter. When $h=5$ the narrow channel and unconfined model curves are nearly overlapping each other, indicating that at this channel width the boundary effects are \emph{almost} negligible. These observations can also be made by looking at the graph of $g_h$ in \figref{fig:extreme}, in particular by considering the difference between $g_h$ at a given value of $h$  and the extreme values at $h=0$, $g_0 = 2$, and at $h= \infty$, $g_\infty = 4$.  

\subsection{Stationary solutions in periodic channels}\label{sec:steady}

In this section we study the steady states of our continuum model and
use the narrow-channel equation as an extension to the one-dimensional
linear ratchet model for Brownian motors. For convenience, we focus on
the one-dimensional equation  \eqref{reduced_conc} in terms of
the coefficient  $g_h$ for the narrow channel cases (NC2) and (NC3)
only. As in the time-dependent case, the numerical solutions of the
PDE are compared with corresponding stochastic simulations of the
individual-based model \eqref{sde}. We suppose that the force along
the channel axis $f_1 (\hat x)$ is the gradient of a potential $V(\hat
x)$, so that  $f_1 (\hat x) = - V'(\hat x)$, where the prime indicates
differentiation. Then \eqref{reduced_conc} can be written as 
\begin{equation}
\label{gradflownarrow}
\frac{\partial \hat p_e}{\partial t}  +
\frac{\partial}{\partial \hat x} \! \left( \hat p_e u  \right)
=0,\quad \mbox{ where }\quad u =- \frac{\partial}{\partial
  x}\big(\log \hat p_e + g_h \phi \hat p_e + V(\hat x)\big). 
\end{equation}
The quantity $u$ can be interpreted as a flow down the gradient of the
free energy $\mathcal F$ \citep{Carrillo:2003uk} associated with
\eqref{reduced_conc}; see \cite{Bruna:2012ub} for more details. The
stationary solution of \eqref{gradflownarrow} with no-flux boundary
conditions is obtained by minimising the free energy, which
corresponds to solving 
\[
\log \hat p_e + g_h \phi \hat p_e + V(\hat x) = C,
\]
with the constant $C$ determined by the normalisation condition on
$\hat p_e$. For the application to ratchet systems, we are interested
in periodic solutions of \eqref{gradflownarrow} with a (constant) flux
$J_0 \equiv \hat p_e u$.\footnote{For periodic boundary data, $J_0$ is
  an extra degree of freedom determined by imposing periodicity.} 

The one-dimensional Fokker--Planck equation \eqref{ntotal} for point
particles with $f_1(\hat x) = - V'(\hat x)$ can be used to model
directed particle transport under \emph{ratchet potentials} (i.e.
potentials spatially asymmetric with respect to their maxima
\citealp{Slater:1997ty}). These potentials may describe a periodic
asymmetric free-energy substrate in the case for molecular motors
through microtubules \citep{Kolomeisky:2007ff} or steric interactions
in the case of polyelectrolytes \citep{Slater:1997ty}. In particular,
theoretical approaches to molecular motors such as kinesin have
focused on either a one-dimensional continuum model such as
\eqref{ntotal} (thus ignoring the interactions between different
motors and the other dimensions) or on stochastic simple-exclusion
models on a lattice \citep{Kolomeisky:2007ff}. More generally, in
these applications the modelling has focused on the form of the
ratchet potential in order to make the model more realistic, instead
on the possible interactions between the particles involved in the
transport. In the remainder of this section we will explore the
effects that interactions (specifically, excluded-volume interactions,
but this could be extended to other types of interactions) can have on
such models. To this end, we consider a specific ratchet model with a
ratchet potential consisting of a periodic part plus a constant
external force or \emph{tilt}. We use the tilted
Smoluchowski--Feynman potential
\begin{equation}
\label{periodicpotential}
V(\hat x, F_0) = \sin(2\pi \hat x) + 0.25 \sin(4 \pi \hat x) - F_0 \hat x.
\end{equation}
Two plots of $V(\hat x)$ for different values of the tilt $F_0$ are shown in \figref{fig:V_F0-1}. 
It can be shown that in the long-time limit the sign of the particle current (or net motion) agrees with the sign of the tilt $F_0$ and that, in the absence of tilt ($F_0 = 0$) there is no net motion \citep{Reimann:2002hs}. An interesting feature of this model is that the relationship between the tilt $F_0$ and the flux $J_0$ is nonlinear (see thick black line in \figref{fig:J0vsF0b}).
\def \scc {1.0}
\def \scl {1.0}
\begin{figure}[htb]
\unitlength=1cm
\begin{center}
\psfragscanon
\psfrag{x}[][][\scl]{$\hat x$}  \psfrag{p}[r][][\scl][-90]{$p_e(x)$} 
\psfrag{V}[][][\scc][-90]{$V(\hat x)\ \ $} 
\psfrag{F0}[][][\scc]{$F_0$} 
\includegraphics[height = .3\linewidth]{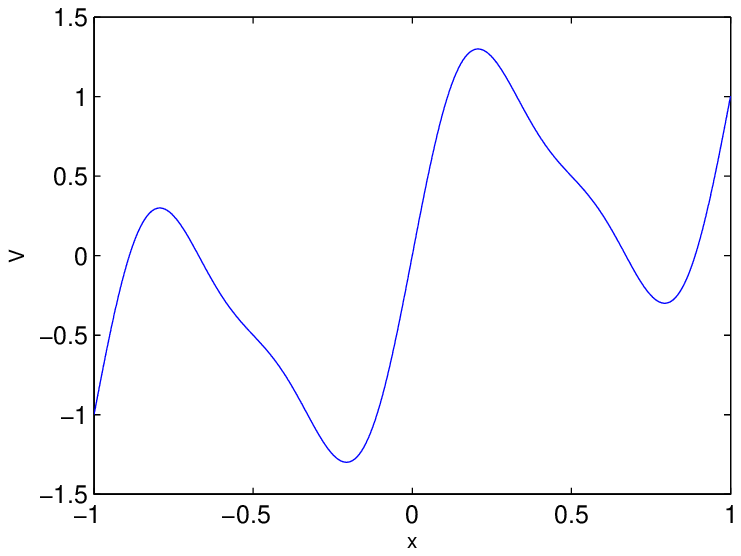} \qquad \includegraphics[height = .3\linewidth]{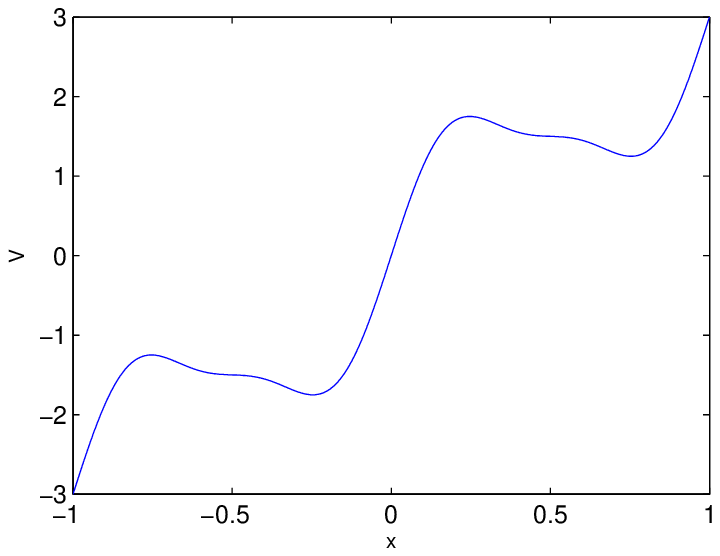}
\caption{Tilted Smoluchowski--Feynman ratchet potential $V(\hat x, F_0)$ in Eq.  \eqref{periodicpotential} for $F_0 = -1$ (left) and $F_0 = -3$ (right).}
\label{fig:V_F0-1}
\end{center}
\end{figure}

\begin{figure}[t!]
\unitlength=1cm
\begin{center}
\psfragscanon
\psfrag{x}[][][\scl]{$\hat x$}  \psfrag{p}[][][\scl][-90]{$\hat p_e \ $} 
\psfrag{F0}[][][\scc]{$F_0$} 
\includegraphics[height = .32\linewidth]{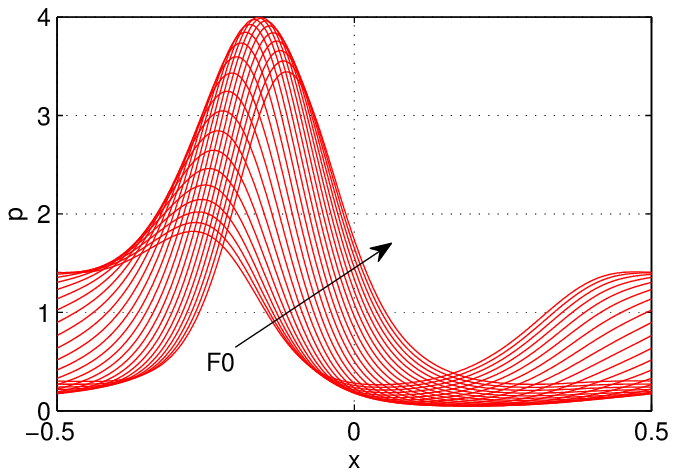} \quad
\includegraphics[height = .32\linewidth]{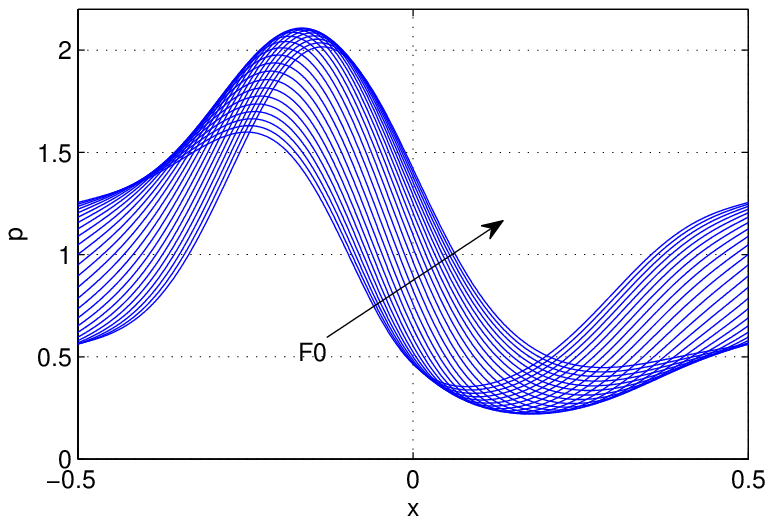}
\caption{Stationary solution $\hat p_e(\hat x)$ under the tilted Smoluchowski--Feynman potential \eqref{periodicpotential} for various values of $F_0$. Stationary solutions of \eqref{gradflownarrow} for point particles, $g_h=0$ (left) and for finite-size particles with $g_h \phi =0.6$ (right).}
\label{fig:p_x_variousF0}
\end{center}
\end{figure}

Next we examine how the $F_0$ vs. $J_0$ relationship changes when nonlinear effects (due to the finite-size of particles and confinement) are included in the model. To this end, we solve for the stationary states of \eqref{gradflownarrow} when $V(\hat x)$ is given by \eqref{periodicpotential} for various tilts $F_0$ and compute the stationary flux $J_0$ of the resulting solution. 
For each tilt $F_0$, we must find the flux $J_0$ and the stationary solution $\hat p_e (\hat x)$ such that the periodic and normalisation conditions are fulfilled:
\begin{align}
\label{steady_ratchet}
\begin{aligned}
(1 + g_h \phi \, \hat p_e) {\hat p_e}' + V(\hat x, F_0) \, \hat p_e &= - J_0,\\
\hat p_e (-1/2) & = \hat p_e(1/2),\\
\textstyle \int_{-1/2}^{1/2} \hat p_e \, \ud \hat x &= 1.
\end{aligned}
\end{align}
 We solve this problem numerically using Chebfun \citep{chebfunv4} in
 MATLAB. Solutions of \eqref{steady_ratchet} for an increasing value
 of the constant force $F_0$ (varying from $-6$ to $+6$) are shown in
 \figref{fig:p_x_variousF0}. The left panel corresponds to point
 particles ($g_h = 0$), while  the right panel corresponds to
 finite-size particles with $g_h \phi = 0.6$.  
The diagram of the resulting steady flux $J_0$ versus the tilt $F_0$
is shown in \figref{fig:J0vsF0b} for increasing values of $g_h \phi
$. We observe that the relationship is nonlinear for point particles
($g_h = 0$, thick black line), but it appears to become linear as
excluded-volume effects get larger (i.e. as $g_h \phi $
increases). 
This is physically reasonable, since point particles get easily
trapped in the wells of the potential, even if these wells are relatively
small. In contrast, it is easier for finite-size particles to escape,
as they may not all fit in the potential well and the nonlinear
diffusion makes the potential barrier easier to overcome. 

\def \scc {1}
\def \scl {1}
\begin{figure}[bt]
\unitlength=1cm
\begin{center}
\psfragscanon
\psfrag{a}[][][\scl]{$g_h \phi$}  
\psfrag{J0}[][][\scc][-90]{$J_0$} 
\psfrag{F0}[t][][\scc]{$F_0$} 
\includegraphics[width = .55\linewidth]{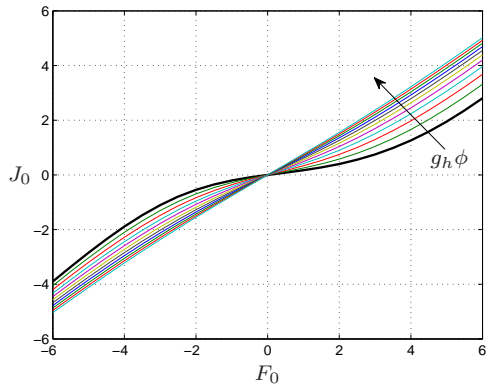}
\caption{Steady-state flux $J_0$ versus tilt $F_0$ from solving \eqref{gradflownarrow} for increasing values of $g_h \phi $ (from 0 to 1). The coloured lines correspond to finite-size particles ($g_h>0$). The thick black line corresponds to point particles ($g_h=0$) as shown in \citet[Figure~2.4]{Reimann:2002hs}.}
\label{fig:J0vsF0b}
\end{center}
\end{figure}

\def \scc {.7} 
\def \scl {1.0}
\begin{figure}[ht]
\unitlength=1cm
\begin{center}
\psfragscanon
\psfrag{x}[][][\scl]{$\hat x$}  \psfrag{u}[][][\scl][-90]{$\hat p_e \ $} 
\psfrag{u0-u015}[r][][\scl]{diff} 
\psfrag{aaaaa0F25}[][][\scc]{$g_h=0, F_0= 2.5$} 
\psfrag{aaaaa015F25}[][][\scc]{$g_h \phi =0.15, F_0= 2.5$} 
\psfrag{aaaaa0Fm6}[][][\scc]{$g_h=0, F_0= -6$} 
\psfrag{aaaaa015Fm6}[][][\scc]{$g_h \phi =0.15, F_0= -6$} 
\hspace{.5cm} \includegraphics[height = .4\linewidth]{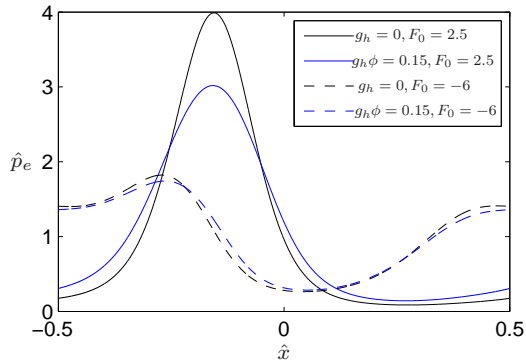} 
\caption{Stationary solution $\hat p_e(\hat x_1)$ under the  potential \eqref{periodicpotential}. Solutions of \eqref{gradflownarrow} for $g_h = 0$ and $g_h=0.15$, and $F_0 = -6$ and $F_0 = 2.5$. }
\label{fig:optimalcase}
\end{center}
\end{figure}

In \figref{fig:optimalcase} we compare the stationary solutions $\hat
p_e$ for point particles  ($g_h = 0$) and finite-size particles (with
an excluded-volume coefficient of $g_h = 0.15$) in two tilting
scenarios: for $F_0 = -6$ and for $F_0 = 2.5$. We observe that while
the solutions are almost overlapping for a tilt of $F_0 = -6$, they
are considerably different for $F_0 =2.5$. This is because for $F_0 =
-6$ the potential $V$ is so tilted that it ceases to have a local
minimum within each period. As a result, the ``advantage'' of
finite-size particles that could more easily overcome the local minima
in the potential is lost.  

We conclude this section by comparing the equilibrium solutions of the
continuum model with the corresponding stochastic simulations of the
discrete model. We use  the Metropolis--Hastings algorithm
\citep{Chib:1995ud} to sample from the stationary density of the
full-particle system, and compare the resulting histogram (averaged
over the cross section) with the stationary solutions $\hat p_e$ of
\eqref{gradflownarrow}. We consider the (NC2) case for which the
coefficient $g_h$ is maximised for a fixed volume fraction $\phi$,
that is, a channel of width $h^* = 1.47$. The other parameters used in
the simulations are $\epsilon = 10^{-3}$, $N = 133$ and $F_0 = 2.5$.  
\def \scc {.7}
\def \scl {1.1}
\begin{figure}[htb]
\unitlength=1cm
\begin{center}
\psfragscanon
\psfrag{x}[][][\scl]{$\hat x$}  \psfrag{y}[r][][\scl][-90]{$\hat y$} 
\psfrag{-h}[][][\scc]{$-h/2 \ \ $} \psfrag{h}[][][\scc]{$h/2 \ \ $} 
\psfrag{0}[][][\scc]{$0$} \psfrag{-0.5}[][][\scc]{$-0.5$} \psfrag{0.5}[][][\scc]{$0.5$} \psfrag{1}[][][\scc]{$1$} \psfrag{1.5}[][][\scc]{$1.5$} \psfrag{2}[][][\scc]{$2$} \psfrag{2.5}[][][\scc]{$2.5$}
\includegraphics[width = .7\linewidth]{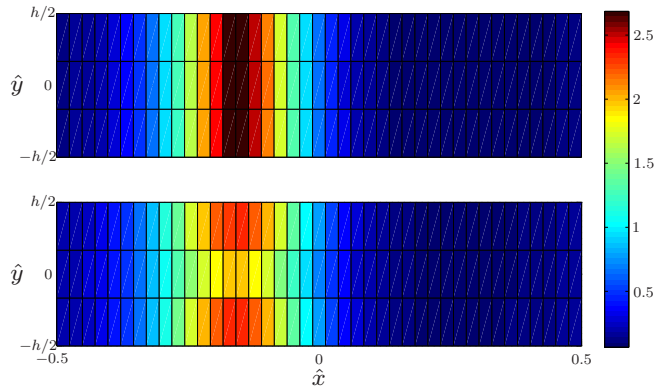}
\caption{Histogram of the stationary density $\hat p(\hat {\bf x})$ under the potential $V(\hat x, F_0)$ in \eqref{periodicpotential} for point particles (top plot) and finite-size particles (bottom plot). Parameters are $F_0 = 2.5$, $h = 1.47$, $\epsilon = 10^{-3}$, $N = 133$. Histograms computed by $10^7$ MH steps.}
\label{fig:stat_a0a1F25_hist}
\end{center}
\end{figure}
\figref{fig:stat_a0a1F25_hist} displays the histograms after $10^7$ steps of the MH algorithm. 
The histogram for point particles (upper plot in \figref{fig:stat_a0a1F25_hist}) does not vary in the cross-sectional direction, as expected. In contrast, the histogram for finite-size particles does display some variation in the $\hat y$-direction: more particles want to be near the boundaries than in the centre of the channel. This is because a hard-disk particle on the boundary excludes only half of the area that would exclude in the centre of the domain (recall that the channel of width $h$ is the domain available to the particles \emph{centres}, not the physical domain), and this is entropically favourable. In our derivation of equation \eqref{fp_reduced} this variation is taken into account but integrated out to obtain the one-dimensional equation. Accounting for the variation in $\hat y$ is not the main objective of this paper but this could be done either by numerically solving the two-dimensional equation \eqref{fp_inta} or evaluating higher order terms in the expansion of section \ref{sec:step2}.

\begin{figure}[htb]
\unitlength=1cm
\begin{center}
\psfragscanon
\psfrag{x}[][][\scl]{$\hat x$}  \psfrag{p}[][][\scl][-90]{$\hat p_e \ $} 
\includegraphics[width = .7\linewidth]{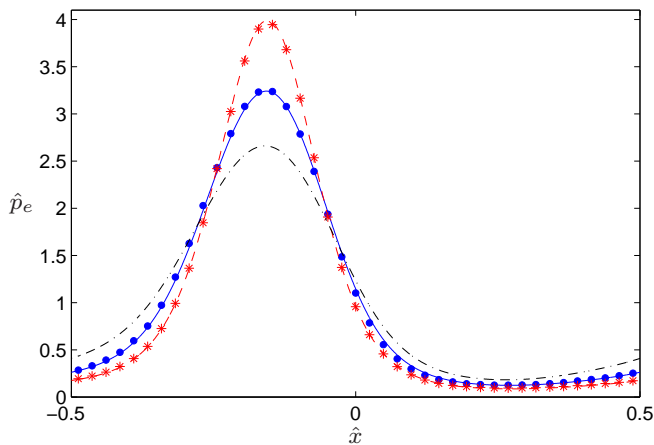}
\caption{One-dimensional stationary density $\hat p_e(\hat x)$ under the potential $V(\hat x, F_0)$ in \eqref{periodicpotential}. Solutions of \eqref{gradflownarrow} for point particles ($g_h=0$, dash red line) and finite-size particles ($g_h=0.1$, solid blue line), with $F_0 = 2.5$, $N = 133$, $h = 1.47$, $\epsilon = 10^{-3}$. Cross-sectional averages of the histograms in  \figref{fig:stat_a0a1F25_hist} for point particles (red asterisks) and finite-size particles (blue circles). Stationary solution of the single-file equation \eqref{p_singlefilelimit} is shown in a dot-dash black line.}
\label{fig:stat_a0a1F25}
\end{center}
\end{figure}

We plot the theoretical predictions by solving \eqref{gradflownarrow}  for both point and finite-size particles alongside their respective simulation counterparts, and observe an excellent agreement in both cases. We examine the importance of taking into account the actual width of the channel, which in this example is only  $h=1.47$, by solving the analogous stationary problem for the single-file model \eqref{p_singlefilelimit}. We plot the result (dot-dash black line in \figref{fig:stat_a0a1F25}) and observe that the single-file model overestimates the excluded-volume effects, as demonstrated by the flatter density profile. 

\section{Discussion} \label{conclu}

In this paper, we have presented a derivation of a continuum model for
the diffusion of finite-size particles in a confined domain whose
dimensions are comparable to the particle dimensions. We have given
the model explicitly for three confined geometries, namely a two- and
three-dimensional narrow channel of square cross-section and
Hele--Shaw cell (two close parallel plates), but also indicated how
the model derivation for more general geometries can be done. The
resulting continuum model is a nonlinear drift-diffusion equation for
the one-particle probability density or for the population volume
concentration, with the nonlinear term depending on the
excluded-volume created by the particles as well as the confinement
parameter (in the cases we have studied, the confinement parameter is
the channel width or the separation between plates). This equation is
defined in the effective domain, a lower-dimensional domain for the
unconfined dimensions only, exploiting the fact that equilibration in
the confined dimensions is relatively fast and most dynamics occur
along the unconfined ones. For example, in the case of a two- or
three-dimensional channel the resulting continuum model is a
one-dimensional equation along the axial direction.  

The derivation of the final continuum model involved two key
steps. First, as in our previous works
\citep{Bruna:2012wu,Bruna:2012cg}, we used the method of matched
asymptotic expansions to reduce the high-dimensional Fokker--Planck
equation associated with the individual-based description of the
system to a low-dimensional Fokker--Planck equation for the
one-particle density function. Second, we exploited the confined
geometry of the domain to perform a further reduction and integrate
out the confined dimensions of the problem. While in our previous
works the particle--particle--wall interactions were a higher-order
correction, and thus were neglected, in this work they were taken into
account. In other words, while in an unconfined domain such three-body
interactions simply create a boundary layer near the domain walls, in
the situations considered in the present work this boundary layer
extends across the cross section and must be solved accurately.  

The model has two interesting features. First, we found that, for a
given volume fraction, there exists an optimal ratio $h$ between
confinement and particle size such that the excluded-volume effects
are maximised. This means, for example, that one can design a
lab-on-a-chip device to achieve a maximal collective diffusion
coefficient. Second, the model is capable of describing the whole
range of confinement levels and interpolating between confinement
extremes, that is, between extreme confinement ($h=0$) when particles
cannot pass each other to unconfined diffusion ($h = \infty$). We have
examined the limiting models corresponding to each of the three
geometries we have studied, and found that our model agrees with
them. For example, in the narrow-channel cases, the extreme
confinement corresponds to a single-file diffusion model
\citep{Rost:1984ts} while the other limit is an unconfined two or
three-dimensional diffusion model \citep{Bruna:2012cg}. This is a
useful analytical tool to predict the error that is being committed by
using the limiting models, that is, either ignoring the fact that
particles can (just) pass each other and using the purely
one-dimensional single-file equation, or neglecting confinement
conditions and boundary layers.  

In order to assess the validity of our model predictions, we have
compared the numerical results of our continuum model for the
two-dimensional narrow channel to the results from the corresponding
individual-based model as well as the limiting continuum models of
point particles, single-file diffusion and unconfined diffusion. We
have observed excellent agreement between the narrow channel model and
the stochastic simulations of the discrete model, and confirmed the
interpolating properties of the model between confinement
extremes. Finally, we have discussed a case study involving the
diffusion under a ratchet potential, used for example to describe the
transport of molecular motors through microtubules. We have examined
the effects that interactions between particles and confinement
conditions can have in the analysis of the problem. For example, we
have found that an increase on excluded-volume effects causes the
particle transport to be less sensitive to the tilting of the ratchet
potential.  

There are several directions along which one can extend this
model. Firstly, it would be interesting to allow multiple types of
particles (labelled blues and reds, say) by combining the analysis in
\cite{Bruna:2012wu} with the 
present work. A 
challenging aspect of this problem is that the resulting model of two
species in a narrow channel must account for the fact that as the
channel becomes single file the order of particles is fixed, and
since the two types of particles are distinguishable the order of particles
matters.
 In other words, we expect a
qualitative difference in the model as $h$ crosses the value of one: for
$h>1$, even if all the red particles were initially to the left of the
blue particles, we expect the two populations to mix together (that
is, the jump in the initial density profiles will spread); in
contrast, for $h<1$, if the two populations are segregated initially,
they will stay like that for all times, and we expect to see this
refelected in the population-level densities. In connection
with this issue is the fact that the self-diffusion of a particle is
not defined in one-dimensional systems \citep{Ackerson:1982ti}.  
Another extension is to allow channels of varying cross section, that
is, to extend the Fick--Jacobs model \eqref{fickjacobs} to the case of
finite-size particles. A simplification of this problem (ignoring the
particle-particle interactions) has been considered by
\cite{Riefler:2010by}. A related problem would be to match a narrow
channel with two bulk domains in each end; such a geometry could have
important applications in the area of ion channels.  

\begin{acknowledgements}
This publication was based on work supported in part by Award
No. KUK-C1-013-04, made by King Abdullah University of Science and
Technology (KAUST). MB acknowledges financial support from EPSRC. We
are grateful to the organisers of the workshop "Stochastic Modelling
of Reaction-Diffusion Processes in Biology", which has led to this
Special Issue.  
\end{acknowledgements}

\appendix 

\section{Derivation of the narrow-channel equation \eqref{fp_reduced} for the (NC2) case}

This section is devoted to the derivation of \eqref{fp_reduced} in the two-dimensional channel (NC2) case. The derivation of the three-dimensional cases (NC3) and (PP) or other (simple) geometries follows similarly (see Subsection \ref{sec:summary} for an outline of the conditions/calculations to be carried out). 

\subsection{Transformation $\mathcal T_1$: Reduction from the individual- to the population-level} \label{sec:step1}

Our starting point is the Fokker--Planck equation for $N$ hard-disc
particles \eqref{fp_eq}, defined in the high-dimensional
(configuration) space $\Omega_\epsilon ^N \subset \mathbb
R^{2N}$. Recall that the configuration space has holes that correspond
to illegal configurations (with particles' overlaps) on which the
no-flux boundary conditions \eqref{fp_bc} hold. The aim of this
subsection is to derive the corresponding Fokker--Planck equation for
the one-particle density $p({\bf x}_1, t) = \int P(\vec x, t)  \ud
{\bf x}_2 \dots {\bf x}_N$, where $P(\vec x, t)$ is the joint
probability density of the $N$ particles. We first note that the
integration of \eqref{fp} over ${\bf x}_2, \dots, {\bf x}_N$ results
in integrals over \emph{contact surfaces} on which $P$ must be
evaluated \citep{Bruna:2012cg}. When the particle volume is small, the
dominant contributions to these contact integrals correspond to
two-particle interactions, so that we can set $N=2$ and then extend
the result to $N$ arbitrary in a straightforward manner. However, in
contrast with the unconfined case studied in \cite{Bruna:2012cg},
under confinement conditions the particle-particle-wall interactions
(three-body) are not negligible and must be taken into account.  
For two particles at positions ${\bf  x}_1$ and ${\bf x}_2$, \eqref{fp} reads 
\begin{subequations}
\label{fp2} 
\begin{align}
\label{fp2_eq}
\frac{\partial P}{\partial t}({\bf  x}_1, {\bf  x}_2, t)
&=  \bo \nabla_{{\bf  x}_1} \cdot \left[  \bo \nabla_{{\bf  x}_1}   P   - {\bf  f}({\bf  x}_1)  P\right] + \bo \nabla_{{\bf  x}_2} \cdot \left[ \bo \nabla_{{\bf  x}_2}   P   - {\bf  f}({\bf  x}_2)  P\right] \quad \text{in} \quad  \Omega_\epsilon ^2,\\
\label{fp2_bc2}
0 &= \left[ \bo \nabla_{{\bf  x}_1}   P -  {\bf  f}(
  {\bf x}_1)  P  \right] \cdot \bo {\hat {\bf n}}_1 +  \left[
  \bo \nabla_{{\bf  x}_2}   P -  {\bf  f}(
  {\bf x}_2)  P \right] \cdot  \bo {\hat {\bf n}}_2, 
\end{align}
\end{subequations}
on ${\bf  x}_i \in \partial \Omega$ and $\|{\bf  x}_1 -
{\bf  x}_2\| = \epsilon$. Here $ \boldsymbol {\hat {\bf n}}_i =  { {\bf n}}_i / \| { {\bf n}}_i\|$, where $ { {\bf n}}_i$ is the
component of the normal vector $\vec  n$ corresponding to the $i$th
particle, $\vec n = ({\bf  n}_1, {\bf  n}_2)$. We note
that $\boldsymbol  {\hat {\bf n}}_1 = 0$ on ${\bf  x}_2 \in \partial
\Omega$, and that $\boldsymbol {\hat {\bf n}}_1 = -\boldsymbol {\hat {\bf n }}_2$ on $\| {\bf x}_1-{\bf  x}_2\| = \epsilon$.

\subsubsection{From $N$ particles to $1$
  particle} \label{sec:step11} 

\begin{figure}[b!]
\begin{center}
\resizebox{0.85\textwidth}{!}{\input{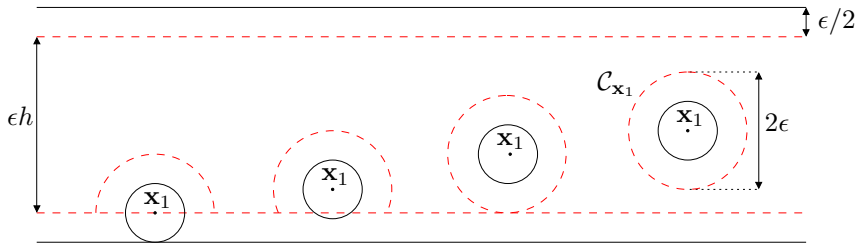}}
\caption{Sketch of the original channel geometry (solid black lines)
  and the effective configuration space for the centre of a second
   circular particle, given by the boundary function $\partial
  \Omega({{\bf x}_1})$ (dash red lines) which depends on the position
  of the first particle ${\bf x}_1$. The
  collision boundary 
  function $\mathcal C_{{\bf x}_1}$ forms part of the effective boundary
  $\partial \Omega({{\bf x}_1})$.} 
\label{fig:moving}
\end{center}
\end{figure}
We denote by $\Omega({\bf  x}_1) =  \Omega \setminus  B_\epsilon
  ({\bf   x}_1)  $ the region available to the
centre of particle 2 when particle 1 is at ${\bf  x}_1$. Note that when the
distance between ${\bf  x}_1$  and $\partial \Omega$ is less than
$\epsilon$ the area $|\Omega({\bf x}_1)|$ increases
(see \figref{fig:moving}) because the area \mbox{$\mathcal  U({\bf x}_1) = B_\epsilon ({\bf  x}_1)\cap\Omega$} excluded by
particle 1 changes with ${\bf x}_1$. The points on which the two
particles are in contact are given by the \emph{collision boundary} 
\begin{equation}
\label{Cx1outer}
\mathcal C_{{\bf x}_1}  = \left \{ {\bf x}_2 \in \Omega \text{ s.t. } \| {\bf x}_2 - {\bf x}_1\| = \epsilon \right \}.
\end{equation}

Integrating Eq. \eqref{fp2_eq} over $\Omega ({\bf  x}_1)$ using the Reynolds transport theorem (on the moving boundary $\mathcal C_{{\bf x}_1}$), the divergence theorem, and the boundary condition \eqref{fp2_bc2} yields 
\begin{align}
\frac{\partial p}{\partial t}({\bf  x}_1, t) = 
{\bo \nabla}_{{\bf  x}_1} \cdot \left[ {\boldsymbol \nabla}_{{\bf  x}_1} \, p -  {\bf  f}({\bf  x}_1) \, p \right] +  \int_{\mathcal C_{{\bf  x}_1}} \! \! - \left ( {\bo \nabla}_{ {\bf x}_1} P + {\bo \nabla}_{ {\bf x}_2} P  \right ) \cdot  {\bf n}_2 \, \mathrm{d} S_2. 
\end{align}
We denote the \emph{collision integral} above by
$\mathcal I$. If we now consider the case of $N$ particles we would
obtain a collision integral for each pair, so that after some particle
relabelling the corresponding equation is
\begin{align}
\label{fp2rr2}
\frac{\partial p}{\partial t}({\bf  x}_1, t) = 
{\bo \nabla}_{{\bf  x}_1} \cdot \left[ {\boldsymbol \nabla}_{{\bf  x}_1} \, p -  {\bf  f}({\bf  x}_1) \, p \right] + (N-1) \! \! \int_{\mathcal C_{{\bf  x}_1}} \! \! - \left ( {\bo \nabla}_{ {\bf x}_1} P + {\bo \nabla}_{ {\bf x}_2} P  \right ) \cdot  {\bf n}_2 \, \mathrm{d} S_2. 
\end{align}

Equation \eqref{fp2rr2} is halfway through transformation $\mathcal
T_1$ (cf.  \figref{fig:scheme}), since the first half of the equation
depends only on $\bf x_1$ while the integral $\mathcal I$ still
depends on the two-particle density $P$ near the collision surface
$\mathcal C_{{\bf x}_1}$.  

At this stage, it is common to use a closure approximation such as
$P({\bf x}_1,{\bf x}_2,t) = p({\bf x}_1,t) p({\bf x}_2,t)$ to evaluate
$\mathcal I$ and obtained a closed equation for $p$
\citep{Rubinstein:1989gk}. However, the pairwise particle
interaction---and therefore the correlation between their
positions---is \emph{exactly} localised near the collision surface
$\mathcal C_{{\bf  x}_1}$. Instead, in the next section we will use an
alternative method based on matched asymptotic expansions to evaluate
$\mathcal I$ systematically \citep{Bruna:2012cg}.  

\subsubsection{Matched asymptotic expansions} \label{sec:step12}

We suppose that when two particles are far apart ($|x_1 -  x_2| \gg
1$) they are independent (at leading order), whereas when they are
close to each other ($| x_1 -  x_2| \sim \epsilon$) they are
correlated. We denote these two regions of the configuration space
$\Omega_\epsilon^2$ the outer region and the inner region,
respectively. We use the $x$-coordinate to distinguish between the
two regions  because the inner region spans 
the channel's cross section. Importantly this implies that the
outer region is disconnected.

\paragraph{Outer region} \label{sec:outer}

In the outer region we consider the change to the narrow-domain
variables \eqref{narrowvariables} and define $\hat P(\hat {\bf x}_1,
\hat{\bf x}_2,t) = \epsilon^{2} P ({\bf  x}_1, {\bf  x}_2, t)$. This
scaling is consistent with that introduced for the one-particle
density in \secref{sec:points}, and is such that $P$ and $\hat P$
each integrate to one in their respective domains $\Omega_\epsilon^2$ and
the narrow-domain variable equivalent $\omega_\epsilon^2$. Then
\eqref{fp2_eq} becomes  
\begin{subequations}
\label{fp3} 
\begin{align}
\label{fp3_eq}
\begin{aligned}
\textstyle \epsilon^2 \frac{\partial \hat P}{\partial t}(\hat {\bf  x}_1, \hat{\bf  x}_2, t)
& =  \textstyle \frac{\partial}{\partial \hat y_1}  \!  \left ( \! \frac{\partial \hat P}{\partial \hat y_1} -  \epsilon f_2(\hat {x}_1, \epsilon \hat y_1) \hat P \right ) + \frac{\partial}{\partial \hat y_2}  \!  \left (  \!  \frac{\partial \hat P}{\partial \hat y_2} - \epsilon f_2(\hat { x}_2, \epsilon \hat y_2) \hat P \right )\\
& \quad \, \textstyle + \epsilon^2   \!  \frac{\partial}{\partial \hat x_1}  \!   \!  \left (  \!  \frac{\partial \hat P}{\partial \hat x_1}  -  f_1(\hat {x}_1, \epsilon \hat y_1) \hat P   \right )+ \epsilon^2  \!   \frac{\partial}{\partial \hat x_2}  \!  \!   \left (  \!  \frac{\partial \hat P}{\partial \hat x_2}  -  f_1(\hat { x}_2, \epsilon \hat y_2) \hat P    \right) \! , 
\end{aligned}
\end{align}
for  $(\hat {\bf  x}_1, \hat {\bf  x}_2) \in \omega_\epsilon ^2$, with 
\begin{alignat}{3}
\label{bc3a}
\textstyle \frac{\partial \hat P}{\partial \hat x_i}  -  f_1(\hat{x}_i, \epsilon \hat y_i) \hat P &= 0& \qquad &\text{on}& \qquad \hat x_i &= \pm \tfrac{1}{2}, \\
\label{bc3b}
\textstyle\frac{\partial \hat P}{\partial \hat y_i}  - \epsilon f_2(\hat{x}_i, \epsilon \hat y_i) \hat P &= 0& \qquad &\text{on}& \qquad \hat y_i &= \pm \tfrac{h}{2},
\end{alignat}
for $i=1,2$.
The boundary condition on the collision line $\mathcal C_{{\bf  x}_1}$
disappears for  \mbox{$|\hat x_1 - \hat x_2| > \epsilon$}, so that it is
``invisible'' to the outer region.\footnote{To see this, we write
  \eqref{Cx1outer} in terms of the  narrow variables. It becomes
  $\mathcal C_{\hat {\bf x}_1} = \{ \hat {\bf x}_2 \in \omega \ : \
  (\hat x_2 - \hat x_1)^2 + \epsilon^2(\hat y_2 - \hat y_1)^2
  =\epsilon^2 \}$.} 
\end{subequations}

In the outer region, we define $P_{out}(\hat {\bf  x}_1, \hat{\bf
  x}_2,t) = \hat P(\hat {\bf  x}_1, \hat{\bf  x}_2,t)$ and look for an
asymptotic solution to \eqref{fp3} by expanding $P_{out}$ in powers of
$\epsilon$.  We find at leading order that  $P_{out}$ must be
independent of the vertical coordinates $\hat y_1$ and $\hat y_2$.
By independence in the outer region we suppose that the leading-order
solution is
separable, 
so
that it is of the form $q(\hat x_1,t) q(\hat x_2,t)$ for some function
$q$. Solving for the next two orders in $\epsilon$ we find that the
solution in the outer region is, to $\mathcal O(\epsilon^2)$, 
\begin{align}
\label{Poutercomplete}
\begin{aligned}
P_{out} (\hat {\bf x}_1  &, \,  \hat {\bf x}_2, t) =  q(\hat x_1,t) q(\hat x_2,t) \\
& + \epsilon \left\{ q(\hat x_1,t) q(\hat x_2,t)  \big[\hat y_1 f_2(\hat {x}_1, 0) + \hat y_2 f_2(\hat {x}_2, 0) \big] + \Upsilon_1(\hat x_1, \hat x_2,t) \right\} \\
& \textstyle +  \epsilon^2 \Big\{ \frac{1}{2}  \left[ \hat y_1^2 \frac{\partial f_2}{\partial y}(\hat x_1, 0) + \hat y_2^2 \frac{\partial f_2}{\partial y}(\hat x_2, 0) + \big ( \hat y_1 f_2(\hat x_1,0)  + \hat y_2 f_2(\hat x_2,0) \big)^2 \right ]  \\
& \phantom{+  \epsilon^2 \Big\{} +  \Upsilon_1(\hat x_1, \hat x_2,t)
\big[ \hat y_1 f_2 (\hat x_1,0) + \hat y_2 f_2(\hat x_2,0) \big]  +
\Upsilon_2(\hat x_1,\hat x_2,t) \Big \}, 
\end{aligned}
\end{align}
where the $\Upsilon_i(\hat x_1, \hat x_2, t)$ are arbitrary functions of
integration, determined by solvability conditions at higher
order. Note that the invariance of $P$ with respect to a switch of
particle labels means that in the outer region both particles have the
\emph{same} density function $q$. From the solvability condition on
the $O(\epsilon^2)$ terms above we 
obtain the following equation for $q$:
\begin{equation}
\label{qdiffeqn}
 \frac{\partial q}{\partial t} (\hat x, t) = \frac{\partial }{\partial \hat x} \left ( \frac{\partial q}{\partial \hat x} - f_1(\hat x,0) q \right).
\end{equation}

\paragraph{Inner region} \label{sec:innerregion}

In the inner region we introduce the \emph{inner variables}
\begin{equation}
\label{inner_vars}
x_1 = \tilde x_1, \qquad y_1= \epsilon \tilde y_1,\qquad 
x_2 = \tilde x_1 + \epsilon \tilde x, \qquad  y_2= \epsilon \tilde y_2,
\end{equation}
and define $\tilde P(\tilde {\bf x}_1, \tilde {\bf x}_2,t) = \epsilon^{2} P ({\bf  x}_1, {\bf  x}_2, t)$. The contact boundary $\mathcal C_{{\bf x}_1}$ \eqref{Cx1outer} becomes
\begin{equation}
\label{Cx1inner}
\tilde {\mathcal C}_{\tilde y_1}  = \Big \{ (\tilde x, \tilde y_2) \in \mathbb R \times [-h/2, h/2]  \ \text{ s.t. } \ \tilde x^2 + (\tilde y_2 - \tilde y_1)^2   = 1 \Big \},
\end{equation}
and problem \eqref{fp2} is transformed to 
\begin{subequations}
\label{fp4} 
\begin{align}
\label{fp4_eq}
\begin{aligned}
\textstyle \epsilon^2 \frac{\partial \tilde P}{\partial t}(\tilde {\bf  x}_1, \tilde {\bf  x}_2, t)
=\ & \textstyle  2 \frac{\partial^2 \tilde P}{\partial \tilde x^2} +  \frac{\partial^2 \tilde P}{\partial  \tilde y_1^2}  +   \frac{\partial^2 \tilde P}{\partial  \tilde y_2^2} - 2\epsilon   \frac{\partial^2 \tilde P}{\partial \tilde x_1  \partial \tilde x} - \epsilon \frac{\partial}{\partial \tilde y_2} \! \left[ f_2(\tilde x_1 + \epsilon \tilde x, \epsilon \tilde y_2) \tilde P \right]    \\
&  \textstyle + \epsilon \frac{\partial}{\partial \tilde x} \! \left \{ \left[ f_1(\tilde x_1, \epsilon \tilde y_1) - f_1(\tilde x_1 + \epsilon \tilde x, \epsilon \tilde y_2) \right] \tilde P \right\}    \\
& \textstyle  -  \epsilon \frac{\partial}{\partial \tilde y_1} \! \left[ f_2(\tilde x_1, \epsilon \tilde y_1) \tilde P \right] +  \epsilon^2  \frac{\partial^2 \tilde P}{\partial  \tilde x_1^2} - \epsilon^2 \frac{\partial}{\partial \tilde x_1} \! \left[ f_1(\tilde x_1, \epsilon \tilde y_1) \tilde P \right],
\end{aligned}
\end{align}
with 
\begin{align}
\label{bcinner}
\begin{aligned}
\textstyle 2 \tilde x \frac{\partial \tilde P}{\partial \tilde x}  + (\tilde y_2 -\tilde y_1) \! \left ( \! \frac{\partial \tilde P}{\partial \tilde y_2} - \frac{\partial \tilde P}{\partial \tilde y_1} \! \right )  &= \textstyle \epsilon \tilde x    \frac{\partial \tilde P}{\partial \tilde x_1} + \epsilon \tilde x  \!\left[ f_1(\tilde x_1\! +\!  \epsilon \tilde x, \epsilon \tilde y_2) - f_1(\tilde x_1 , \epsilon \tilde y_1) \right] \! \tilde P \\
 & \ \ \, +\epsilon (\tilde y_2 - \tilde y_1) \! \big[  f_2 (\tilde x_1 \! + \! \epsilon \tilde x, \epsilon \tilde y_2) - f_2(\tilde x_1, \epsilon \tilde y_1) \big] \! \tilde P,
\end{aligned}
\end{align}
on $\tilde {\mathcal C}_{\tilde y_1}$ and
\begin{alignat}{3}
\label{nfinner1}
\textstyle \frac{\partial \tilde P}{\partial \tilde y_1} &=\epsilon f_2(\tilde x_1, \epsilon \tilde y_1) \tilde P & \qquad &\text{on}& \qquad \tilde y_1 &= \textstyle \pm \frac{h}{2} ,\\
\label{nfinner2}
\textstyle \frac{\partial \tilde P}{\partial \tilde y_2} &=\epsilon f_2(\tilde x_1 + \epsilon \tilde x, \epsilon \tilde y_2) \tilde P & \qquad &\text{on}& \qquad \tilde y_2 &= \textstyle \pm \frac{h}{2}.
\end{alignat}
In addition to \eqref{bcinner}-\eqref{nfinner2}, the inner solution must match with the outer as $\tilde x \to \pm \infty$. Expanding the outer solution in terms of the inner variables, which corresponds to replacing $\hat x_1 = \tilde x_1$,  $\hat y_1 = \tilde y_1$,  $\hat x_2 = \tilde x_1 + \epsilon \tilde x$ and $\hat y_2 = \tilde y_2$ in \eqref{Poutercomplete}, and subsequently expanding in powers of $\epsilon$, we obtain the following \emph{matching condition}:
\begin{align}
\label{match}
\begin{aligned}
\tilde P & \sim \textstyle q ^2  + \epsilon \left [ \tilde x q     \frac{\partial q}{\partial \tilde x_1}  + (\tilde y_1 + \tilde y_2) f_2 q^2(\tilde x_1) + \Upsilon_1(\tilde x_1, \tilde x_1) \right] \\
& \textstyle \quad + \epsilon^2  \Big [  \frac{\tilde x^2}{2}  q  \frac{\partial^2  q}{\partial \tilde x_1^2 } +   \tilde x (\tilde y_1 + \tilde y_2) f_2 q  q_{\tilde x_1} + \tilde x \tilde y_2 \frac{\partial f_2}{\partial x}q^2 + \tilde x   \frac{\partial \Upsilon_1}{\partial \hat x_2} (\tilde x_1,\tilde x_1)\\
& \textstyle \phantom{\quad + \epsilon^2  \Big [} + \frac{1}{2} \left( (\tilde y_1^2  + \tilde y_2^2)\frac{\partial f_2}{\partial y} + (\tilde y_1   + \tilde y_2)^2  f_2^2 \right) q^2 + (\tilde y_1  + \tilde y_2) f_2  \Upsilon_1(\tilde x_1, \tilde x_1) \\
&  \phantom{ \quad + \epsilon^2  \Big [} + \Upsilon_2(\tilde x_1,\tilde x_1) \Big] + \cdots \quad \qquad \text{as} \qquad \tilde x \to \pm \infty,
\end{aligned}
\end{align}
\end{subequations}
where $ q \equiv q(\tilde x_1, t)$. 
Expanding $\tilde P$ in powers of $\epsilon$, $\tilde P\sim  \tilde P^{(0)} + \epsilon \tilde P^{(1)}  + \epsilon^2 \tilde P^{(2)} + \cdots$, we find that the leading- and first-order solutions of  \eqref{fp4} are
\begin{align}
\label{solo0}
\tilde P^{(0)} &= q^2 (\tilde x_1),\\
\label{solo1}
\tilde P^{(1)} &= \tilde x \, q (\tilde { x}_1)   \frac{\partial q  }{\partial \tilde x_1} (\tilde x_1) +(\tilde y_1 + \tilde y_2)f_2(\tilde x_1, 0) q^2(\tilde x_1) + \Upsilon_1(\tilde x_1, \tilde x_1) .
\end{align}
Unlike in the bulk or unconfined problem \citep{Bruna:2012wu,Bruna:2012cg}, the narrow-channel problem requires computing the second-order inner solution. The solution procedure is rather cumbersome and is omitted here. It involves a further change of variable $\tilde x = \sqrt 2 \tilde s$ to turn the problem into a Poisson problem, and solving two sub-problems numerically using the commercial finite-element solver COMSOL Multiphysics 4.3. The second-order solution of \eqref{fp4} is (see Appendix C.1 in \citealp{Bruna:2012ub} for full details)
\begin{align}
\label{solfinalP2}
\begin{aligned}
\tilde P^{(2)}   = \  & \textstyle  \frac{1}{2}  \tilde x^2 q   \frac{\partial^2  q}{\partial \tilde x_1^2 } + \frac{1}{2} \! \left (\tilde y_1^2 + \tilde y_2^2 \right) q^2 \!  \left( \frac{\partial f_2}{\partial y} + f_2^2 \right) + \tilde y_1 \tilde y_2 f_2^2 q^2  \\
& \textstyle +  \tilde x \left ( \tilde y_1  f_2 q   \frac{\partial q}{\partial \tilde x_1} +  \tilde y_2  q   \frac{\partial  (f_2 q) }{\partial \tilde x_1} \right )  + \left(\frac{\partial f_1}{\partial y} -  \frac{\partial f_2}{\partial x} \right) q^2 \tilde Q_2(\tilde x, \tilde y_1, \tilde y_2) \\
&\textstyle   + \sqrt 2 \Big  [q  \frac{\partial^2  q}{\partial  \tilde x_1^2 } -  \big( \frac{\partial q}{\partial  \tilde x_1}\big) ^2  - \frac{\partial f_1}{\partial x} q^2 \Big]   \tilde Q_1(\tilde x, \tilde y_1, \tilde y_2) + \Upsilon_2(\tilde x_1, \tilde x_1),
\end{aligned}
\end{align}
with  $\tilde Q_i (\tilde x, \tilde y_1, \tilde y_2) = \tilde v_i (\tilde x/\sqrt 2 , \tilde y_1, \tilde y_2)$, where $\tilde v_1(\tilde s, \tilde y_1, \tilde y_2)$ and $\tilde v_2 (\tilde s, \tilde y_1, \tilde y_2)$ are given by (numerical solutions of)
\begin{alignat}{3}
\nonumber
\widetilde {\bo \nabla}^2  \tilde v_1  &= 0, & & & &\\
\label{probforv1}
\widetilde {\bo \nabla} \tilde v_1 \cdot \tilde {\bo \nu}  &  =  \tilde s^2 & \qquad  &\text{on} &\qquad &\tilde {\mathcal  D}_{\tilde y_1},\\
\nonumber
\widetilde {\bo \nabla} \tilde v_1 \cdot \tilde {\bo \nu}  &= 0 & &\text{on} & & \tilde y_i=\pm \frac{h}{2},\\ 
\nonumber
\tilde v_1 &\sim D_1 |\tilde s|&  &\text{as} & & \tilde s \to \pm \infty,
\end{alignat}
and 
\begin{alignat}{3}
\nonumber
\widetilde {\bo \nabla}^2  \tilde v_2  &= 0, & & & &\\
\label{probv2}
\widetilde {\bo \nabla} \tilde v_2 \cdot \tilde {\bo \nu}  &  =  \tilde s (\tilde y_1 - \tilde y_2), & \qquad  &\text{on} &\qquad &\tilde {\mathcal  D}_{\tilde y_1},\\
\nonumber
\widetilde {\bo \nabla} \tilde v_2 \cdot \tilde {\bo \nu}  &= 0, & &\text{on} & & \tilde y_i=\pm \frac{h}{2},\\ 
\nonumber
\tilde v_2 &\sim 0 ,&  &\text{as} & & \tilde s \to \pm \infty.
\end{alignat}
Here $\widetilde {\bo \nabla}$ stands for the gradient operator with respect to the position vector $(\tilde s, \tilde y_1, \tilde y_2)$, $\tilde {\mathcal D}_{\tilde y_1} $ is the transformed collision boundary $\tilde {\mathcal C}_{\tilde y_1}$  \eqref{Cx1inner}, and $\tilde {\bo \nu}$ is the  outward unit normal on this mapped boundary, $\tilde {\bo \nu} = - \frac{\sqrt 2}{2} (2 \tilde s, \tilde y_1 - \tilde y_2, \tilde y_2 - \tilde y_1)$. Finally, the constant field $D_1$ at infinity for $\tilde v_1$ is related to the integration function from the outer $\Upsilon_1$, 
\begin{equation*}
D_1 = \frac{\lim_{\hat x_2 \to \hat x_1} \frac{\partial \Upsilon_1}{\partial \hat x_2} (\hat x_1, \hat x_2)}{ \left [ q \frac{\partial^2 q}{\partial  \tilde x_1^2} - \big (\frac{\partial q}{\partial  \tilde x_1} \big ) ^2  - \frac{\partial f_1}{\partial x} q^2 \right] }.
\end{equation*}
Out of the derivation we also find that $\Upsilon_1(\hat x_1, \hat
x_2) \equiv \Upsilon_1(|\hat x_1-\hat x_2|)$ with $\Upsilon_1(x)$
differentiable satisfying $\Upsilon_1(0) = 0$. In contrast, the
contribution from the other outer function $\Upsilon_2$ is left
unknown (it could be determined by matching higher order terms), but
we are able to ignore it as it has a zero contribution to the
collision integral $\mathcal I$ (as we will see in the next section). We
note that, in  \eqref{solfinalP2}, $q$, $f_1$ and $f_2$ are functions
of the ``outer'' variable $\tilde x_1$ only, namely, $q = q(\tilde
x_1, t)$ and $f_i = f_i(\tilde x_1, 0)$. 

Combining \eqref{solo0}, \eqref{solo1} and \eqref{solfinalP2} we have the solution to the inner problem \eqref{fp4} up to $\mathcal O(\epsilon^2)$. 

\subsubsection{Collision integral} \label{sec:step13}

Now we go back to Eq. \eqref{fp2rr2} and use the asymptotic
solution of the previous subsection to turn it into an equation for
$p({\bf x}_1, t)$ only thus completing transformation $\mathcal
T_1$. Note that, since the integral $\mathcal I$ is over the collision
boundary $\mathcal C_{{\bf x}_1}$, it lives in the inner region and we
must use $\tilde P$ to evaluate it.

\begin{figure}[htb]
\begin{center}
\resizebox{0.5\textwidth}{!}{\input{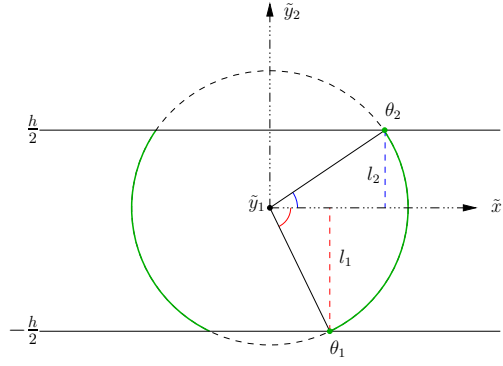}}
\caption{Domain of integration $\tilde {\mathcal C}_{\tilde y_1}$ (solid green line), distances $l_1$ and $l_2$ between the vertical coordinate of the first particle and the lower and upper channel walls and corresponding angles $\theta_1$ and $\theta_2$.}
\label{fig:line_integration}
\end{center}
\end{figure}

In terms of the inner variables, $\mathcal I$ is
\begin{align}
\label{integral_inner}
 \mathcal I = \epsilon^{-2} \int_{\tilde {\mathcal C}_{\tilde y_1}}  \textstyle \left[ (\tilde y_2- \tilde y_1) \! \left ( \frac{\partial \tilde P}{\partial \tilde y_1} + \frac{\partial  \tilde P}{\partial \tilde y_2} \right )   +\epsilon   \tilde x\frac{\partial  \tilde P}{\partial \tilde x_1} \right] \ud \tilde l,
\end{align}
where $\tilde {\mathcal C}_{\tilde y_1}$ is given in \eqref{Cx1inner} and $\ud \tilde l$ is the line integral along this curve (for $\tilde y_1$ fixed, see \figref{fig:line_integration}). 
Depending on the channel width $h$ (relative to one, which is the
radius of $\tilde {\mathcal C}_{\tilde y_1}$), the integration is over
the whole circle or a part of it. Introducing the distances $l_1 =
\max(-h/2-\tilde y_1,-1)$ and $l_2 = \min(h/2-\tilde y_1, 1)$, the
angles at contact with the lower and upper channel walls are
$\theta_1 = \arcsin l_1$ and $\theta_2 = \arcsin l_2$,
respectively. (These are equal to $\pm \pi/2$ for no contact.) 

Writing $\mathcal I = \epsilon^{-2} \big(  \mathcal I^{(0)} +  \epsilon \mathcal I^{(1)} + \epsilon^2\mathcal I^{(2)} + \cdots \big)$ and using \eqref{integral_inner}, we find that
\begin{align}
\label{Iinner0}
&\mathcal I^{(0)}  =  0,\\
\label{Iinner-1}
& \mathcal I^{(1)} = 
2f_2 q^2 \mu_0(h, \tilde y_1),\\
\label{finalmentI2}
& \! \begin{aligned}
\mathcal I^{(2)} = \ & \textstyle 2q^2 \left( \frac{\partial f_2}{\partial y} + 2 f_2^2 \right) \tilde y_1 \mu_0(h, \tilde y_1) + \left (2q  \frac{\partial ^2 q}{\partial \tilde x_1^2} - q^2 \frac{\partial f_1}{\partial x} \right) \mu_1(h, \tilde y_1)  \\
& \textstyle +  q^2 \left( \frac{\partial f_2}{\partial y} + 2f_2^2 \right)  \mu_2(h, \tilde y_1)  + 2 \left(\frac{\partial f_1}{\partial y} -  \frac{\partial f_2}{\partial x} \right) q^2 \mathcal J[\tilde Q_2] (h, \tilde y_1)  \\
& \textstyle + 2 \sqrt 2 \Big[ q \frac{\partial ^2 q}{\partial \tilde x_1^2} - \big ( \frac{\partial  q}{\partial \tilde x_1} \big)^2  - \frac{\partial f_1}{\partial x} q^2 \Big] \mathcal J[\tilde Q_1](h, \tilde y_1) ,
\end{aligned}
\end{align}
where $f_i = f_i (\tilde x_1, 0)$,
\begin{subequations}
\label{functionmus}
\begin{align}
\label{functionmu0}
\mu_0(h, \tilde y_1) &= \int_{\tilde {\mathcal C}_{\tilde y_1}} \! (\tilde  y_2 - \tilde y_1) \, \ud \tilde l =  \textstyle 2  \big(\sqrt{1-l_1^2}  - \sqrt{1-l_2^2} \big),\\
\label{functionmu1}
\mu_1(h, \tilde y_1) &= \int_{\tilde {\mathcal C}_{\tilde y_1}} \! \tilde x^2 \, \ud \tilde l = \textstyle l_2 \sqrt{1-l_2^2} - l_1 \sqrt{1-l_1^2} + \arcsin l_2 - \arcsin l_1,\\
\label{functionmu2}
\mu_2(h, \tilde y_1) &=  \int_{\tilde {\mathcal C}_{\tilde y_1}} \! (\tilde y_2 - \tilde y_1)^2 \, \ud \tilde l  = \textstyle  l_1 \sqrt{1-l_1^2} -l_2 \sqrt{1-l_2^2}   - \arcsin l_1 + \arcsin l_2,
\end{align}
\end{subequations}
and $\mathcal J$ is the integral operator $\mathcal J[Q](h, \tilde y_1) = \! \int_{\tilde {\mathcal C}_{\tilde y_1}} \!  \left[  Q_{\tilde y_2}  (\tilde y_2- \tilde y_1)  +   Q_{\tilde x} \tilde x  \right]  \ud \tilde l$.  The terms $\mathcal J[\tilde Q_1]$ and $\mathcal J[\tilde Q_2]$  are evaluated numerically with COMSOL (see Appendix C.2 in \citealp{Bruna:2012ub} for more details).

\subsubsection{Population-level Fokker--Planck equation} \label{sec:finalintegrated}

Combining \eqref{Iinner-1} and \eqref{finalmentI2} we obtain the first two terms of the asymptotic expansion for $\mathcal I$, which depends on both the channel width $h$ and the elevation of the first particle $\tilde y_1$ but is independent of the position of the second particle. Thus we can drop the first particle label (the subindex 1) for clarity of notation. Inserting this expansion into \eqref{fp2rr2}, we obtain an equation for the first particle
\begin{align}
\label{fp_inta}
\frac{\partial p}{\partial t}({\bf  x}, t) = 
{\bo \nabla}_{{\bf  x}} \cdot \left[ {\boldsymbol \nabla}_{{\bf  x}} \, p -  {\bf  f}({\bf  x}) \, p \right] + (N-1) \Big(  \epsilon^{-1}  \mathcal I^{(1)} + \mathcal I^{(2)} \Big) \qquad \text{in} \qquad \Omega,
\end{align}
which involves the marginal density $p({\bf x},t)$, the outer density
$q(\hat x,t)$ and the channel width $h$. This concludes the
transformation $\mathcal T_1$  from $N$ particles to one  particle (see \figref{fig:scheme}).

\subsection{Transformation $\mathcal T_2$: Reduction of the number of geometric dimensions} \label{sec:step2}

Following a similar procedure to that for  point particles  in
\secref{sec:points}, we will reduce \eqref{fp_inta} into a
one-dimensional effective equation along the axial direction. First,
integrating \eqref{fp2_bc2} over $\Omega({\bf x})$ for ${\bf x}
\in \partial \Omega$, we obtain the following no-flux boundary
condition: 
\begin{equation}
\label{problempbc}
\left[ {\bo \nabla}_{{\bf  x}}  p- {\bf f}({\bf x}) p \right] \cdot \hat{ \bf n} = 0 \qquad \text{on} \qquad \partial \Omega.
\end{equation}
Analogously to the point-particles case, we use the narrow-domain variables \eqref{narrowvariables} and define $\hat p(\hat {\bf x}, t) = \epsilon p({\bf x}, t)$. With this rescaling, Eqs. \eqref{fp_inta} and \eqref{problempbc} become
\begin{subequations}
\label{problempnarrow}
\begin{align}
\label{fp2r_narrow}
 \epsilon ^2 \frac{\partial \hat p}{\partial t}(\hat {\bf  x}, t) &=   \frac{\partial}{\partial \hat y} \Big(   \frac{\partial \hat p}{\partial \hat y} - \epsilon f_2 \hat p \Big) + \epsilon ^2 \frac{\partial}{\partial \hat x}  \Big( \frac{\partial \hat p}{\partial \hat x} - f_1 \hat p \Big) +  (N-1) \epsilon^2 \big( \mathcal I^{(1)} + \epsilon \mathcal I^{(2)} \big),\\
\label{bconpx}
 \frac{\partial \hat p}{\partial \hat x} &= f_1 \hat p \qquad  \text{on}  \qquad  \hat x = \pm \frac{1}{2},\\ 
\label{bconpy}
 \frac{\partial \hat p}{\partial \hat y} &=  \epsilon f_2 \hat p \qquad  \text{on}  \qquad  \hat y = \pm \frac{h}{2},
\end{align}
\end{subequations}
where $f_i \equiv f_i (\hat x, \epsilon \hat y)$. There is no need to expand the integral terms $\mathcal I^{(i)}$ in terms of the narrow-domain variables since these are written in terms of the inner variables \eqref{inner_vars}, which coincide with the narrow-domain variable for expressions independent of the second particle's coordinates. 

Expanding $\hat p$ in powers of $\epsilon$ and solving \eqref{problempnarrow} gives, at leading order, that $\hat p$ is independent of  $\hat y$. As before, we introduce the effective one-dimensional densities as $\hat p_e^{(i)} = \int_{-h/2} ^{h/2} \hat p^{(i)} \ud \hat y$. Thus we have that $\hat p_e ^{(0)} \equiv h \hat p_e^{(0)}$. At the next order,  
\begin{align}
\hat p^{(1)}(\hat {\bf x}, t)  &= f_2(\hat x, 0) \hat p^{(0)}(\hat x, t) \, \hat y + \hat p^{(1)}_e (\hat x,t) / h.
\end{align}
For clarity of notation, in the remaining of this section we write $f_i(\hat x, 0) \equiv f_i$. Integrating the second order of \eqref{fp2r_narrow} over the channel's cross section and using \eqref{bconpy}, gives
\begin{align}
\label{p0total}
\frac{\partial \hat p_e^{(0)}}{\partial t}(\hat x, t) &=  \frac{\partial}{\partial \hat x} \left( \frac{\partial \hat p_e^{(0)}}{\partial \hat x} - f_1 \hat p_e^{(0)} \right),
\end{align}
where we have used that $ \int_{-h/2}^{h/2} \mathcal I^{(1)} \ud \hat y = 0$ [see Eqs. \eqref{Iinner-1} and \eqref{functionmu0}]. Note that this equation coincides with the effective equation for point particles \eqref{ntotal}. It is at the next order that the finite-size effects appear. 

Repeating the same procedure of integrating with respect to $\hat y$ the $\mathcal O(\epsilon^3)$ of \eqref{problempnarrow} and using \eqref{bconpy} to eliminate $\hat p^{(3)}$ yields the following \emph{solvability condition}
\begin{equation}
\label{p1solve}
\frac{\partial \hat p^{(1)}_e}{\partial t}(\hat x, t)   -\frac{\partial}{\partial \hat x} \left( \frac{\partial \hat p^{(1)}_e}{\partial \hat x} - f_1 \, \hat p^{(1)}_e  \right)  = (N-1)  \int _{-h/2}^{h/2}  \mathcal I^{(2)} \, \ud \hat y.
\end{equation}
Using \eqref{finalmentI2} and \eqref{functionmus}, the cross-section integral of $\mathcal I^{(2)}$ is
\begin{align}
\label{ptsolve2}
\begin{aligned}
\int _{-h/2}^{h/2}  \mathcal I^{(2)} \ud \hat y = \ &  \textstyle  \left (2q  \frac{\partial ^2 q}{\partial \tilde x_1^2} - q^2 \frac{\partial f_1}{\partial x} \right) M_1(h) + 2 \left(\frac{\partial f_1}{\partial y} -  \frac{\partial f_2}{\partial x} \right) q^2 \mathcal M[\tilde Q_2]  \\
&\textstyle +2 \sqrt 2 \Big[ q \frac{\partial ^2 q}{\partial \tilde x_1^2} - \big ( \frac{\partial  q}{\partial \tilde x_1} \big)^2  - \frac{\partial f_1}{\partial x} q^2 \Big] \mathcal M[\tilde Q_1],
\end{aligned}
\end{align}
where $M_1 (h) = \int_{-h/2}^{h/2} \mu_1(h, \hat y) \ud \hat y$ reads
\begin{align}
\label{functionM1}
M_1(h) =   \pi h - \frac{4}{3}  + \Theta(1-h) \left[ \frac{2}{3} (2 + h^2) \sqrt{1-h^2}  - 2h \arccos(h) \right],
\end{align}
where $\Theta(x)$ is the Heaviside step function, and $\mathcal M [Q](h) = \int _{-h/2}^{h/2} \mathcal J[Q](h, \hat y) \, \ud \hat y$. 
Although $\tilde Q_1$ and $\tilde Q_2$ are only solved numerically, using information from their respective problems \eqref{probforv1} and \eqref{probv2} one can deduce analytical expressions for their integrals $\mathcal M[\tilde Q_i]$, namely that $\mathcal M[\tilde Q_1] = - M_1(h)/(2 \sqrt 2)$ and $\mathcal M[\tilde Q_2] = 0$ (see Appendix C.3 in \citealp{Bruna:2012ub}). Using this, we find that 
\begin{align}
\label{2IaIb}
\int _{-h/2}^{h/2}  \mathcal I^{(2)} \ud \hat y = \frac{\partial}{\partial \tilde x_1} \left( q \frac{\partial q}{\partial \tilde x_1} \right) M_1(h).
\end{align}
Because  $q$ is independent of the inner variables ($\tilde x, \tilde y_1, \tilde y_2$), we can write $q(\tilde x_1, t) = q_e (\hat x, t)/h$. Moreover, the  normalisation condition on $\hat P$ gives that $q_e (\hat x, t) = \hat p_e (\hat x, t) + \mathcal O(\epsilon)$. Therefore the right-hand side of \eqref{2IaIb} becomes $ \frac{ \partial}{ \partial {\hat x} } \big ( \hat p_e \frac{\partial \hat p_e}{\partial {\hat x} } \big )/h^2$. 

Combining \eqref{p0total}, \eqref{p1solve} and \eqref{2IaIb} yields
\begin{align}
\label{pafinal}
\frac{\partial \hat p_e}{\partial t}(\hat x, t)  =  \frac{\partial}{\partial \hat x} \left\{ \left[1 + (N-1) \epsilon \tfrac{M_1(h)}{h^2} \hat p_e \right] \frac{\partial \hat p_e}{\partial \hat x} - f_1(\hat x, 0)  \hat p_e  \right\},
\end{align}
which coincides with the effective equation \eqref{fp_reduced} for a
two-dimensional narrow-channel after writing $\alpha_h \equiv
M_1(h)/h^2$; see \eqref{functiong}.  

\subsection{Outline of steps for other geometries} \label{sec:summary}

In this section we indicate the key steps to derive the effective continuum Fokker--Planck equation \eqref{fp_reduced} for a general geometry, and in particular for the three-dimensional cases (NC3) and (PP) presented in \secref{sec:finitesize}. First, we note below relevant definitions that change with the problem dimension $d$ and the number of confined dimensions $k$ (recall that $d_e = d-k$): 
\begin{flushenumerate}
\item {\em Identify the number of confined and effective dimensions:} the original position vector is split into two components, ${\bf x}=({\bo x}_e, {\bo x}_c)$ (see Table \ref{table:dimensions}). For example, for (NC3) ${\bo x}_e = x$ and ${\bo x}_c = (y,z)$, while for (PP) ${\bo x}_e = (x, y)$ and ${\bo x}_c = z$.
\begin{table}[htp]
\begin{center}
\begin{tabular}{l|c|c|c|}
& Dimension & Variables & Domain \\
Original problem & $d$ & ${\bf x}$ & $\Omega$\\
Confined space & $k$ & ${\bo x}_c$ & $\Omega_c$ \\
Effective problem & $d_e$ & ${\bo x}_e$ & $\Omega_e$
\end{tabular}
\end{center}
\caption{Dimension and variables in each of the spaces and subspaces.}
\label{table:dimensions}
\end{table}%

\item {\em Determine the confinement parameter(s):}
Next we must choose an scaling for the confined dimensions relative to the unconfined ones. In all cases considered here we made that simple by saying that all confined dimensions are of length $H$ relative to the unconfined ones, but there could be of different lengths too, such as the narrow channel of rectangular cross section mentioned briefly in Subsec. \ref{sec:otherlimits}. Suppose that the confinement dimensions are ${\bo H} = (H_1, \dots, H_k) = \mathcal O(\epsilon)$. Then the vector of confinement parameters is given by ${\bo h} = (h_1, \dots, h_k)$, with $h_i = H_i/\epsilon$. 

Note that we are assuming that a confined dimension is always of order $\epsilon$ (the particle's diameter). However, this could also be generalised and introduce an intermediate scaling (between $\epsilon$ and one). 

\item {\em Narrow-domain variables transformation:} the generalisation of the change of variables \eqref{narrowvariables} is
\begin{equation*}
{\bo x}_e = \hat{\bo x}_e, \qquad {\bo x}_c = \epsilon \hat {\bo x}_c.
\end{equation*}
We write $\hat {\bf x} = (\hat {\bo x}_e, \hat {\bo x}_c)$. 
The one-particle and two-particle densities in the rescaled domain are respectively defined as
\begin{align*}
\hat p (\hat{\bf x}_1,t) = \epsilon^{k} p({\bf x}_1,t),\qquad
\hat P (\hat{\bf x}_1, \hat {\bf x}_2,t) = \epsilon^{2k} P ({\bf x}_1,  {\bf x}_2,t).
\end{align*}
The original domain $\Omega$ is mapped into the rescaled domain $\omega$. 
 
\item {\em Apply the effective domain transformation:}
The $\mathcal T_2$ transformation to reduce the original $d$-dimensional problem to a $d_e$-dimensional problem (cf. Subsec. \ref{sec:step2}) requires the following rescaled densities
\begin{align*}
 \hat p_e (\hat x_1, t) = \Lambda \,  \hat p(\hat {\bf  x}_1, t) ,\qquad  \hat P_e (\hat x_1, \hat x_2, t) = \Lambda^2 \hat P(\hat {\bf  x}_1, \hat{\bf  x}_2, t),
\end{align*}
where
\begin{equation}
\textstyle  \Lambda = \prod_{i=1}^k h_i.
\end{equation}
This factor is, in fact, equal to the volume of the rescaled domain $\omega$ (also $|\omega| \equiv |\omega_c|$). Recall it is introduced so that $\hat p_e$ and $\hat P_e$ are defined as densities. 

\item {\em Evaluate the collision integral $\mathcal I$:} 
The evaluation of the contribution of the two-particle interaction $\mathcal I$ reduces to computing one coefficient like $M_1({\bo h})$ in \eqref{pafinal} for each of the unconfined dimensions. Suppose that all the unconfined dimensions are symmetric. In general it is equal to
\begin{equation}
M_1({\bo h}) = \int_{\omega_c} \mu_1({\bo h},\hat {\bo x}_c) \, \ud \hat {\bo x}_c,
\end{equation}
where $\hat {\bo x}_c$ are the confined coordinates of the first particle (it was simply $\hat y$ in the (NC2) [cf. \eqref{functionM1}]. 
The function $\mu_1({\bo h},\hat {\bo x}_c)$ is the integral of $\hat x^2$ over 
the contact surface between two particles when the first one has 
coordinates $(\hat {\bo x}_e, \hat {\bo x}_c)$. Without loss of generality we 
can set $\hat {\bo x}_e \equiv {\bo 0}_e$.  In the rescaled problem, this surface is a 
$d-$dimensional unit sphere centred at $({\bo 0}_e, \hat {\bo x}_c)$, and 
(possibly) intersected with the confinement walls $\partial \omega_c$:
\begin{equation}
\mu_1({\bo h},\hat {\bo x}_c)  = \int_{B(\hat {\bo x}_c) \cap \omega_c} x^2 \,  \ud S,
\end{equation}
where $B(\hat {\bo x}_c)$ is the unit ball, $\ud S$ is the surface differential and $x$ is the unconfined dimension of this surface. Once $M_1$ is computed, we use it for the nonlinear coefficient of the effective Fokker--Planck equation. The generalisation of the coefficient $\alpha_h$ in \eqref{fp_reduced} is given by
\begin{equation}
\alpha_{\bo h} = \frac{1}{\Lambda} M_1({\bo h}).
\end{equation}
\end{flushenumerate}

\bibliographystyle{spbasic}      

\begin{thebibliography}{34}
\providecommand{\natexlab}[1]{#1}
\providecommand{\url}[1]{{#1}}
\providecommand{\urlprefix}{URL }
\expandafter\ifx\csname urlstyle\endcsname\relax
  \providecommand{\doi}[1]{DOI~\discretionary{}{}{}#1}\else
  \providecommand{\doi}{DOI~\discretionary{}{}{}\begingroup
  \urlstyle{rm}\Url}\fi
\providecommand{\eprint}[2][]{\url{#2}}

\bibitem[{Ackerson and Fleishman(1982)}]{Ackerson:1982ti}
Ackerson BJ, Fleishman L (1982) {Correlations for dilute hard core
  suspensions}. J Chem Phys 76:2675--2679.

\bibitem[{Alberts et~al(2002)Alberts, Johnson, Lewis, Raff, Roberts, and
  Walter}]{alberts2002molecular}
Alberts B, Johnson A, Lewis J, Raff M, Roberts K, Walter P (2002) {Molecular
  biology of the cell}. New York: Garland Science.

\bibitem[{Bodnar and Vel{\'a}zquez(2005)}]{Bodnar:2005kv}
Bodnar M, Vel{\'a}zquez JJL (2005) {Derivation of macroscopic equations for
  individual cell-based models: a formal approach}. Math Meth Appl Sci
  28(15):1757--1779.

\bibitem[{Bruna(2012)}]{Bruna:2012ub}
Bruna M (2012) {Excluded-volume Effects in Stochastic Models of Diffusion}. DPhil
  thesis, University of Oxford.

\bibitem[{Bruna and Chapman(2012{\natexlab{a}})}]{Bruna:2012wu}
Bruna M, Chapman SJ (2012{\natexlab{a}}) {Diffusion of multiple species with
  excluded-volume effects}. J Chem Phys 137(20):204,116--204,116--16.

\bibitem[{Bruna and Chapman(2012{\natexlab{b}})}]{Bruna:2012cg}
Bruna M, Chapman SJ (2012{\natexlab{b}}) {Excluded-volume effects in the
  diffusion of hard spheres}. Phys Rev E 85(1):011103.

\bibitem[{Burada et~al(2009)Burada, H{\"a}nggi, Schmid, and
  Talkner}]{Burada:2009hr}
Burada PS, H{\"a}nggi P, Schmid G, Talkner P (2009) {Diffusion in Confined
  Geometries}. ChemPhysChem 10(1):45--54.

\bibitem[{Carrillo et~al(2003)Carrillo, McCann, and Villani}]{Carrillo:2003uk}
Carrillo JA, McCann RJ, Villani C (2003) {Kinetic equilibration rates for
  granular media and related equations: entropy dissipation and mass
  transportation estimates}. Rev Mat Iberoam 19(3):971--1018.

\bibitem[{Chib and Greenberg(1995)}]{Chib:1995ud}
Chib S, Greenberg E (1995) {Understanding the metropolis-hastings algorithm}.
  Am Stat 49:327--335.

\bibitem[{Dekker(2007)}]{Dekker:2007kc}
Dekker C (2007) {Solid-state nanopores}. Nat Nanotechnol 2(4):209--215

\bibitem[{Eichhorn et~al(2002)Eichhorn, Reimann, and
  H{\"a}nggi}]{Eichhorn:2002du}
Eichhorn R, Reimann P, H{\"a}nggi P (2002) {Brownian Motion Exhibiting Absolute
  Negative Mobility}. Phys Rev Lett 88(19):190601.

\bibitem[{Erban et~al(2007)Erban, Chapman, and Maini}]{Erban:2007we}
Erban R, Chapman SJ, Maini PK (2007) {A practical guide to stochastic
  simulations of reaction-diffusion processes}. Arxiv preprint arXiv:07041908.

\bibitem[{H{\"a}nggi and Marchesoni(2009)}]{Hanggi:2009ha}
H{\"a}nggi P, Marchesoni F (2009) {Artificial Brownian motors: Controlling
  transport on the nanoscale}. Rev Mod Phys 81(1):387.

\bibitem[{Henle et~al(2008)Henle, DiDonna, Santangelo, and
  Gopinathan}]{Henle:2008fv}
Henle ML, DiDonna B, Santangelo CD, Gopinathan A (2008) {Diffusion and binding
  of finite-size particles in confined geometries}. Phys Rev E 78(3):031118.

\bibitem[{Hille(2001)}]{Hille:2001tw}
Hille B (2001) {Ion Channels of Excitable Membranes}. Sinauer, Sunderland, MA.

\bibitem[{Howorka and Siwy(2009)}]{Howorka:2009iw}
Howorka S, Siwy Z (2009) {Nanopore analytics: sensing of single molecules}.
  Chem Soc Rev 38(8):2360--2384.

\bibitem[{Jacobs(1967)}]{jacobs1967diffusion}
Jacobs MH (1967) {Diffusion Processes}. Springer-Verlag, New York.

\bibitem[{Keil et~al(2000)Keil, Krishna, and Coppens}]{Keil:2000ev}
Keil FJ, Krishna R, Coppens MO (2000) {Modeling of Diffusion in Zeolites}. Rev
  Chem Eng 16(2):71--197.

\bibitem[{Klumpp et~al(2005)Klumpp, Nieuwenhuizen, and
  Lipowsky}]{Klumpp:2005ed}
Klumpp S, Nieuwenhuizen TM, Lipowsky R (2005) {Movements of molecular motors:
  Ratchets, random walks and traffic phenomena}. Physica E 29(1-2):380--389.

\bibitem[{Kolomeisky and Fisher(2007)}]{Kolomeisky:2007ff}
Kolomeisky AB, Fisher ME (2007) {Molecular Motors: A Theorist's Perspective}.
  Annu Rev Ecol Evol Syst 58(1):675--695.

\bibitem[{Lizana and Ambj{\"o}rnsson(2009)}]{Lizana:2009ic}
Lizana L, Ambj{\"o}rnsson T (2009) {Diffusion of finite-sized hard-core
  interacting particles in a one-dimensional box: Tagged particle dynamics}.
  Phys Rev E 80(5):051103.

\bibitem[{Mu{\~n}oz-Guti{\'e}rrez et~al(2012)Mu{\~n}oz-Guti{\'e}rrez,
  Alvarez-Ram{\'\i}rez, Dagdug, and Espinosa-Paredes}]{MunozGutierrez:2012cz}
Mu{\~n}oz-Guti{\'e}rrez E, Alvarez-Ram{\'\i}rez J, Dagdug L, Espinosa-Paredes G
  (2012) {Diffusion in one-dimensional channels with zero-mean time-periodic
  tilting forces}. J Chem Phys 136(11):114103.

\bibitem[{Nicolau~Jr. et~al(2007)Nicolau~Jr., Hancock, and
  Burrage}]{NicolauJr:2007dy}
Nicolau~Jr DV, Hancock JF, Burrage K (2007) {Sources of anomalous diffusion on
  cell membranes: a Monte Carlo study}. Biophys J 92(6):1975--1987.

\bibitem[{Plank and Simpson(2012)}]{Plank:2012fa}
Plank MJ, Simpson MJ (2012) {Models of collective cell behaviour with crowding
  effects: comparing lattice-based and lattice-free approaches}. J R Soc
  Interface 9(76):2983--2996.

\bibitem[{Pries et~al(1996)Pries, Secomb, and Gaehtgens}]{Pries:1996ik}
Pries AR, Secomb TW, Gaehtgens P (1996) {Biophysical aspects of blood flow in
  the microvasculature}. Cardiovasc Res 32(4):654--667.

\bibitem[{Reguera and Rub{\'\i}(2001)}]{Reguera:2001ev}
Reguera D, Rub{\'\i} J (2001) {Kinetic equations for diffusion in the presence
  of entropic barriers}. Phys Rev E 64(6):061106.

\bibitem[{Reimann(2002)}]{Reimann:2002hs}
Reimann P (2002) {Brownian motors: noisy transport far from equilibrium}. Phys
  Rep 361(2):57--265.

\bibitem[{Riefler et~al(2010)Riefler, Schmid, Burada, and
  H{\"a}nggi}]{Riefler:2010by}
Riefler W, Schmid G, Burada PS, H{\"a}nggi P (2010) {Entropic transport of
  finite size particles}. J Phys: Condens Matter 22(45):454109.

\bibitem[{Rost(1984)}]{Rost:1984ts}
Rost H (1984) {Diffusion de sph\'eres dures dans la droite r\'eelle: Comportement
  macroscopique et equilibre local.} In: Az{\'e}ma J, Yor M (eds) S{\'e}minaire
  de Probabilit{\'e}s XVIII 1982/83, Springer Berlin, pp 127--143.

\bibitem[{Rubinstein and Keller(1989)}]{Rubinstein:1989gk}
Rubinstein J, Keller JB (1989) {Particle distribution functions in
  suspensions}. Phys Fluids A A1(10):1632--1641.

\bibitem[{Scala et~al(2007)Scala, Voigtmann, and De~Michele}]{Scala:2007fd}
Scala A, Voigtmann T, De~Michele C (2007) {Event-driven Brownian dynamics for
  hard spheres}. J Chem Phys 126(13):134109.

\bibitem[{Slater et~al(1997)Slater, Guo, and Nixon}]{Slater:1997ty}
Slater GW, Guo HL, Nixon GI (1997) {Bidirectional transport of polyelectrolytes
  using self-modulating entropic ratchets}. Phys Rev Lett 78(6):1170--1173.

\bibitem[{Trefethen and {others}(2011)}]{chebfunv4}
Trefethen LN, {others} (2011) {Chebfun Version 4.2}. The {C}hebfun
  {D}evelopment {T}eam.

\bibitem[{Zwanzig(1992)}]{Zwanzig:1992ta}
Zwanzig R (1992) {Diffusion past an entropy barrier}. J Phys Chem
  96(10):3926--3930.

\end{thebibliography}

\end{document}